%% file: E-BOSS_AA.final.tex
\documentclass[structabstract]{aa}  
\usepackage{graphicx}
\usepackage{txfonts}
\usepackage{longtable,lscape}
\begin{document}
   \title{E-BOSS: An Extensive stellar BOw Shock Survey. 
I: Methods and First Catalogue}

   \author{C. S. Peri\inst{1,2},
           P. Benaglia\inst{1,2},
           D. P. Brookes\inst{3},
           I. R. Stevens\inst{3},
           \and N. Isequilla\inst{2}  }

   \institute{Instituto Argentino de Radioastronom\'{\i}a, 
CCT-La Plata (CONICET), C.C.5, (1894) Villa Elisa, Argentina\\
              \email{cperi@fcaglp.unlp.edu.ar}
         \and
             Facultad de Ciencias Astron\'omicas y Geof\'isicas, UNLP, 
Paseo del Bosque s/n, (1900) La Plata, Argentina
         \and
             School of Physics and Astronomy, University of Birmingham, 
Edgbaston, Birmingham B15 2TT, UK}

   \date{Received Month number, 2011; accepted Month number, 2011}


\authorrunning{Peri et al.}
\titlerunning{The Extensive stellar BOw Shock Survey} 
  \abstract
   {Bow shocks are produced by many astrophysical objects where shock 
waves are present. Stellar bow shocks, generated by runaway stars, have 
been previously detected in small numbers and well-studied. Along with 
progress in model development and improvements in observing instruments, 
our knowledge of the emission produced by these objects and its origin 
can now be more clearly understood.}
   {We produce a stellar bow-shock catalogue by applying uniform search 
criteria and a systematic search process. This catalogue is a starting 
point for statistical studies, to help us address fundamental questions 
such as, for instance, the conditions under wich a stellar bow shock is 
detectable.}
   {By using the newest infrared data releases, we carried out a search 
for bow shocks produced by early-type runaway stars. We first explored 
whether a set of known IRAS bow shock candidates are visible in the most 
recently available IR data, which has much higher resolution and sensitivity. 
We then carried out a selection of runaway stars from the latest, large 
runaway catalogue available. In this first release, we focused on OB stars 
and searched for bow-shaped features in the vicinity of these stars.}
   {We provide a bow-shock candidate survey that gathers a total of 28 
members which we call the Extensive stellar BOw Shock Survey (E-BOSS). 
We derive the main bow-shock parameters, and present some preliminary 
statistical results on the detected objects.}
   {Our analysis of the initial sample and the newly detected objects 
yields a bow-shock detectability around OB stars of $\sim$ 10 per cent. 
The detections do not seem to depend particularly on either stellar mass, 
age or position. The extension of the E-BOSS sample, with upcoming IR data, 
and by considering, for example, other spectral types as well, will allow 
us to perform a more detailed study of the findings.}


   \maketitle

\section{Introduction}

Many astrophysical objects perturb the interstellar medium and 
produce different kinds of observable structures. Early-type 
stars with large peculiar velocities (termed runaway stars) are 
one example of the perturbing agent. They are responsible for the 
generation of stellar bow shocks.

Runaway stars have been studied for several decades. There are 
currently two proposed mechanisms for the origin of their high 
velocities (Hoogerwerf et al. 2000). One is the binary supernova 
scenario (BSS, Zwicky 1957, Blauuw 1961) and the other is the 
dynamical ejection scenario (DES, Poveda et al. 1967, Gies \& 
Bolton 1986). In the BSS, the runaway star that was originally 
part of a binary system, acquires a high speed when its companion
explodes as a supernova. In the context of the DES, the runaway 
star achieves its velocity thanks to the dynamical interaction 
with one or more stars. In some cases, the trajectory of the stars 
and the original systems can be reconstructed, and the mechanism 
that has kicked the star can be identified. The gathered evidence 
has not always enabled strong conclusions to be made (see for 
instance the dissenting findings by Comer\'on \& Pasquali 2007 
and Gvaramadze \& Bomans 2008). Studies such as Moffat et al. 
(1998) of runaway stars suggest that the BSS is slightly more 
likely, and that those stars with significant peculiar supersonic 
motion relative to the ambient ISM, tend to form bow shocks in the
direction of the motion.

Abundant studies of individual or smaller groups of runaway stars
can be found. However, compilations of large number of high velocity 
stars are scarce. A good example is the Galactic O-Star Catalog (GOSC, 
Ma\'{\i}z Apell\'aniz et al. 2004), which contains the physical 
properties of about 40 of these stars. Tetzlaff et al. (2010) carried 
out an extensive kinematical and probability study, that allowed us to 
identify thousands of runaway stars. The authors used the Hipparcos 
catalogue (Perryman et al. 1997) and built a database of $\sim$ 2500 
objects.

The effects of runaway stars on the environment and the interaction 
between those stars and the interstellar medium (ISM) have been 
widely documented. A fundamental contribution can be found in Van Buren 
and McCray (1988), who presented a list of 15 bow shock-like candidates, 
found while studying Galactic HII regions detected in IRAS (Infrared 
Astronomical Satellite) data. Their study triggered an extended 
search for stellar bow-shock candidates, which resulted in almost 60 
similar sources of which around 20 were actually candidates (Van 
Buren et al. 1995, Noriega-Crespo et al. 1997: NC97). Some of the
candidates were studied by Brown \& Bomans (2005), using the H${\alpha}$ 
all-sky survey (SHASSA/VTSS). They searched images of 37 objects of the 
list of Van Buren et al. (1995) and detected 8 bow shocks. They also 
calculated environmental parameters and in all the cases they found 
consistency with the features of the warm ionized medium.

In the past few years, searches of relatively small sky regions have 
revealed more stellar bow shocks. Povich et al. (2008) reported the 
discovery of six stellar bow shocks in the star-forming regions M~17 
and RCW~49 from Spitzer GLIMPSE (Galactic Legacy Infrared Mid-Plane 
Survey Extraordinaire) images. By combining 2MASS (Two Micron All-Sky 
Survey), Spitzer, MSX (Midcourse Space eXperiment), and IRAS data, 
they obtained the SEDs of the bow shocks and stars associated with 
them. Other 10 bow-shock candidates were found by Kobulnicky et al. 
(2010) using mid-IR images from the Spitzer Space Telescope Cygnus 
X Legacy Survey. Arnal et al. (2011) exposed a study of the surrounding 
emission related to the star HD~192281, a member of the OB association 
Cygnus OB8. They analyzed neutral hydrogen, radio continuum, molecular 
gas (CO) emission and infrared (IR) data, and derived several parameters 
of the medium. They concluded that HD~192281 seems to generate a stellar 
bow shock and it is possible that triggered star formation is underway.
Bow shocks have been seen around some massive X-ray binaries, such as 
Vela X-1 (Kaper et al. 1997) and 4U\,1907+09 (Gvaramadze et al. 2011a).

Gvaramadze et al. (2011b) carried out a search for OB stars running 
away from the young stellar cluster NGC\,6357. They discovered seven 
bow shocks, and thoroughly discussed the scenario for these structures 
to develop.

During the investigation of the outskirts of Cygnus OB2, Comer\'on \& 
Pasquali (2008) discovered a bow-shaped structure close to the high 
mass runaway star BD\,+43\degr\,3654. From spectroscopic data, they 
re-classified the star and derived its stellar mass, ranking the star
as one of the three most massive runaways known in the Milky Way. 
Stellar proper motions as well as MSX observations confirmed that 
the bow shock could be produced by the star, in turn probably a 
former member of the mentioned association. In a subsequent 
publication, Benaglia et al. (2010) carried out a study of the bow 
shock produced by BD\,+43\degr\,3654, and evaluated whether the 
structure could give rise to high energy emission. By means of 
dedicated VLA observations, they calculated the spectral index in 
the region of the bow shock and measured values compatible with a 
non-thermal radiation origin. Assuming that there is a population 
of relativistic particles, they estimated the SEDs generated by 
different radiative processes.

In spite of these studies, numerous questions remain to be answered. 
How common are stellar bow shocks? In which conditions are they 
produced and detected? Questions such as how the spectral type, 
velocity, stellar wind, coordinates of the star, and other parameters 
influence the formation of a bow shock will be more clearly addressed 
as more examples are analyzed.

In this paper, we introduce the Extensive BOw Shock Survey (E-BOSS) 
generated by means of on-line available data from different missions. 
We used the first-release Wide-field Infrared Survey Explorer (WISE) 
images, as a tool to detect new structures and to contribute to what 
is known about those already studied.

In the next section, we briefly discuss the theoretical background 
and analyze the conditions that could increase the probability of 
detecting a stellar bow shock. Section 3 describes the procedure 
followed to obtain the sample in which we search for bow-shock 
candidates. In Section 4, we present the bow shocks found and 
characterize them. Section 5 shows the statistics performed, 
and Sect. 6 has a corresponding discussion. In the last section, 
we comment on the immediate prospects.


\section{Theoretical considerations}

Runaway stars move through the interstellar medium (ISM) with 
velocities that overcome the field stars velocities and the 
sound of speed in the ISM. They shock the ISM and produce the 
so-called bow shocks that sweep up the ISM matter into thin, 
dense shells (Wilkin 1996). The theoretical study of bow shocks 
has been developed not only from an analytical point of view but 
also with numerical simulations.
 
Wilkin (1996) derived analytical exact solutions for stellar wind 
bow shocks in the thin-shell limit, stressing the importance of the 
conserved momentum within the shell. He developed a simple method 
to reproduce the  shape of the shell, mass column, and velocity of 
the shocked gas throughout the shell. Later, Wilkin (2000) improved 
the model studying the modifications of the bow shocks in two 
cases where: (i) a star moves supersonically with respect to an 
ambient medium with a density gradient perpendicular to the stellar 
velocity, and (ii) a star with a mis-aligned, axisymmetric wind 
moves in a uniform medium. He found that the region of the stand-off 
point (where ambient and wind pressures balance each other) is tilted 
in both cases. In that way, the star does not lie in the line that 
divides the bow shock into two halves.

Dgani et al. (1996a, 1996b) peformed a stability analysis of thin 
isothermal bow shocks. In the first paper, they showed that the 
bow shocks produced by stars with fast stellar winds are more stable 
than those generated by slow winds. In the second article, the 
authors then investigated non-linear instabilities and run numerical 
simulations to solve the problem. They proposed that the slower the 
wind, the highest the instability. They also applied the model to 
the star $\alpha$ Cam and concluded that the clumps observed might 
be explained by the instabilities.

Comer\'on \& Kaper (1998) conducted a semi-analytical study of 
bow shocks produced by OB runaway stars and derived expressions 
that suggest different results from those obtained with the 
assumption of instantaneous cooling of the shocked gas (Wilkin 1996). 
They ran numerical simulations that reveal a wealth of details in 
the formation, structure, and evolution of the bow shocks. These 
features strongly depend on the conditions of the medium and star. 
The bow shocks can either form or not; and if they do form, they 
can be either stable, unstable, or layered.

To complete the study of bow shocks and improve the models, they
need to be observed and analyzed. In principle, owing to the 
presence of the shocks that the runaway stars produce in the ISM, 
the dust is heated and in that way re-radiates at infrared 
wavelengths. As we describe in other sections, bow shocks have 
been observed at infrared (e.g. van Buren et al. 1995, NC97), 
optical (e.g. Brown \& Bomans 2005), and -in a few cases- radio
wavelengths, and will eventually be detected at high energy waves
(Benaglia et al. 2010).


\section{The making of the E-BOSS sample}

We searched for stellar bow shocks in various different databases 
described later. We account for a number of criteria that help us 
to identify new stellar bow shocks.

\begin{enumerate}
\item We examined the surroundings of early-type runaway stars, as 
these have high velocities and strong winds that sweep up interstellar 
matter and also high luminosities that contribute to the dust heating. 
\item We selected nearby stars ($d <$ 3 kpc), which thus have brighter
bow shocks. 
\end{enumerate}

We generated the initial sample in two different ways. We considered 
first bow shock candidates that had previously been detected in the 
IRAS data (NC97). Secondly, we carried out a systematic search around 
runaway stars that could produce bow shocks, using the catalogue of 
Tetzlaff et al. (2010). The next section describes the process in 
more detail. In the rest of the paper, we divide the E-BOSS sample 
into two groups, groups 1 and 2.

{\subsection{Group 1}

For this group, we took the bow shock candidates database of NC97. 
These authors searched for bow shocks and other features at IR 
wavelengths from the HiReS-IRAS maps. Table 1 lists the 56 objects 
that are OB stars from NC97, hereafter called group 1 (the WR stars
HD 50896 and HD 192163, from the original list, were excluded).

{\subsection{Group~2} 

The second group was extracted from the catalogue of Tetzlaff et al. 
(2010). From a sample of 7663 young stars observed by Hipparcos 
(Perryman et al. 1997), the authors built a catalogue of 2546 runaway 
stars candidates. Their study lists stellar names, velocities, spectral
types, ages, and masses. From this last list, we found 244 stars of 
spectral type O to B2. We list this 244 stars in Table 2; they comprise 
what we call hereafter group~2. A total of 17 stars are common to both 
groups.

\subsection{Information collation}

We used the NASA/IPAC Infrared Science 
Archive\footnote{http://irsa.ipac.caltech.edu/} to gather relevant
infrared and sub-millimeter missions data (IRAS, MSX, WISE, Spitzer, 
etc). In practice, data from MSX and 
WISE\footnote{http://wise.ssl.berkeley.edu/} 
have proven to be the most useful. In particular, WISE has 
discovered various bow shocks. Although the published WISE data 
covers half of the sky, in effect, it covers more than two thirds 
of both group 1 and group~2 samples. When further data becomes 
available, it will be added to a subsequent release of the E-BOSS 
sample.

In addition to the infrared data, we searched for bow shock emission 
at H$\alpha$, using the Virginia Tech. Spectral Survey (VTSS, Dennison 
et al. 1997) and the Southern Hemispheric H$\alpha$ Sky Survey Atlas 
(SHASSA, Gaustad et al. 2001). These surveys cover most of the sky 
with a spatial resolution of 0.8\arcmin\ for SHASSA and 1.6\arcmin\ 
for VTSS. We did not find any convincing bow shock candidates for 
either our group~1 or group~2 sample. This is in contrast to Brown 
\& Bomans (2005), who found 8 possible bow shock candidates, starting 
with a target list of 37 stars from van Buren et al. (1995). We note 
that the stars are often located in complex H$\alpha$ emission regions 
making identification of a bow shock feature difficult. In some cases, 
we did detect possible H$\alpha$ emission from a bow shock but which 
is not coincident with a clear bow shock detected by WISE (for example, 
HD\,48099 and HD\,149757). In these cases, we relied on the infrared 
detection.

The detection of radio emission and the measurement of a non-thermal 
radio spectral index coincident with the location of the bow-shock 
candidate related to the star BD\,$+$43\degr\,3654 encouraged us to 
check for low frequency radio emission around the E-BOSS members. We 
used the postage stamp server for NRAO/VLA Sky Survey (NVSS, Condon 
et al. 1998), which returns radio images of the sky in various formats. 
Most of the radio maps contain only point sources, which are not
particularly related to the IR features. However, three E-BOSS 
objects, apart from BD\,+43\degr\,3654, correspond interesting 
comma-shaped radio sources at 1.4 GHz. A thorough investigation 
based on dedicated radio observations towards these candidates, 
HIP\,88652, HIP\,38430, and HIP\,11891 is under way (Peri et al., 
in preparation).


\section{Results}

\subsection{Group~1} 

We searched for WISE and MSX data toward the 56 group-1 targets, 
at all available wavelengths, with a field size of 1 sq degree. 
About 55\% of the targets have MSX data, and 70\% have WISE data. 
Table 1 gives the Hipparcos number, alternative names, and the 
spectral type of the stars as in NC97, or updated from GOSC 
(Ma\'{\i}z Apell\'aniz et al. 2004) whenever possible. Columns 4 
and 5 list the results obtained with MSX and WISE. We identified 
several types of structures around the stars that are represented
with different symbols in Table 1. The categories are: a point 
source at the position of the star, diffuse emission in the vicinity 
of the star, no emission, a bow-shaped emission feature, a bubble 
candidate, and extended emission in the field. A '?' symbol implies 
doubtful feature. The last column reproduces the results of NC97. 
Two special cases are noted in the table. Cappa et al. (2008) 
studied the environs of HD 92206, and identified an extended 
HII region. The star HD 36862 is very close to HD 36861 (O8III); 
they both belong to the rich cluster $\lambda$ Ori (Bouy et al. 
2009), are not single isolated objects, and have a large spatial
velocity. In both cases, a bow shock might be hidden and we 
discarded these two objects from our sample.

We found 18 bow-shock candidates (BS-C) out of the 56 stars of 
group~1 (Table 1). Of the 18 BS-C, 3 of them were detected with 
MSX (one is the case of BD +43${^\circ}$ 3654) and the other 15 
with WISE. Out of a total of 18 BS-C, 14 were also classified as 
BS-C by NC97 (see Table 1). A total of 4 new BS-C were found here, 
related to HIP 24575, HIP 38430, HIP 77391, and HIP 114990. We 
believe that the different results between Group~1 stars and NC97 
are due to the high resolution and sensitivity of the more recent 
data.
 
\subsection{Group~2} 

We list the 244 O-B2 runaway candidates extracted from Tetzlaff 
et al. (2010) in Table 2. We searched the WISE data to identify 
BS-C. The table is divided into two parts: the upper part contains 
those stars for which WISE data were released, and the lower 
portion where they were not.

We found a total of 17 BS-C, marked with bold font in the top 
part of Table 2; seven of them had been identified in group~1. 

\subsection{The E-BOSS sample}

The 28 objects of the E-BOSS sample are shown in Figures 1 to 5.
In Table 3, the star name is given in the first column, the group 
membership in column 2, and Galactic coordinates in columns 3 and 
4. Spectral types are in column 5. The distances, in column 6, 
were taken from several sources, namely Megier et al. (2009), 
Mason et al. (1998), Schilbach \& R\"oser (2008), Hanson (2003), 
and Thorburn et al. (2003), or derived from Hipparcos parallaxes 
(van Leeuwen 2007). The wind terminal velocities $v_{\infty}$ are 
either from Howarth et al. (1997) or derived using table 3 of 
Prinja et al. (1990). To compute the stellar mass-loss rates 
$\dot{M}$, we used the routine described by Vink et al. (2001), 
and stellar parameters derived from standard models (e.g. Martins 
et al. 2005). The stellar tangential velocities $v_{\rm tg}$ 
were taken from Tetzlaff et al. (2010), or derived from proper 
motions (van Leeuwen 2007, see last two columns of Table 3). 
Radial velocities $v_{\rm r}$ are from the Second Catalogue of 
Radial Velocities with Astrometric Data (Kharchenko et al. 2007).

In Figs. 1 to 5, we show the WISE and MSX images of the 28 BS-C. 
Superimposed on the images, we have plotted with arrows two directions, 
representing the proper motion of the star derived by van Leeuwen 
(2007), and that of the corrected proper motion after taking into 
account the Galactic rotation of the ISM at the location of the 
star (as done for example in Comer\'on \& Pasquali 2007 and Moffat 
et al. 1998, 1999).

For each BS-C, we measured geometrical parameters, such as the
spatial extent $l$, the width $w$, and the distance from the 
star to the midpoint of the bow shock structure $R$ (Table 4, 
columns 2 to 7).

The ISM ambient density in the vicinity of the star $ n_{\rm ISM}$ 
can be estimated using the expression that gives the so-called 
stagnation radius $R_0$ (see Wilkin 1996 for definition and details)

$$R_0 =  \sqrt{ \frac{ \dot{M} v_\infty }{ 4 \pi \rho_{\rm a} v_*^2} } \, ,$$

\noindent where the ambient medium density is 
$\rho_{\rm a} = \mu \, n_{\rm ISM}$ and $v_*$ is the spatial stellar
 velocity. We estimated the volume density of the ISM in H atoms at 
the bow shock position assuming $R_0 \sim R$, a mass per H atom 
$\mu = 2.3 \times 10^{-24}$ g, and the helium fractional abundance 
$Y = 0.1$. The values obtained for $n_{\rm ISM}$ are given in column 
8 of Table 4, and should be interpreted with caution. In many cases, 
the width of the bow shock is substantial compared to $R$, which adds 
an uncertainty to the $R_0$ values used in the last equation.

There are additional factors that might affect the values of $R_0$ 
and $n_{\rm ISM}$, such as errors in the mass-loss rate due to clumping, 
in addition to the potential source of errors in the parameters used, 
as described above.


\section{Statistics}

In Figs. 6-7, we present the $(l,b)$ distribution of group 1 and 
group 2 stars, showing those with and without bow shocks. There does
not seem to be any preferential location for stars with bow shocks. 
We note that there are no bow shocks at high latitudes, but that the 
small number of stars there means we cannot say anything conclusive.

Figs. 8-9 show the occurrence of bow shocks as a function of spectral
type for each group. We might have expected more detections of bow 
shocks from the more massive, earlier-type, or fastest stars, which 
is not seen. The number of stars in each subtype grows with later 
spectral types, probably reflecting that there is no strong bias 
in the Group~2 stellar sample.

Figs. 10-11 show that bow shocks were detected around both the
lower mass stars and the youngest stars, but the very small numbers
 in many of the bins prevents us to draw any clear conclusion.


\section{Discussion}

The E-BOSS sample introduced here constitutes the most substantial 
sample of stellar bow shocks and should provide a reliable basis 
to more detailed studies of the structure and formation of bow shocks.

For group~1, the availability of MSX and WISE images has enabled us 
to improve previous results, which relied on the IRAS data. Some 
structures were identified as bow shocks, others rejected, and in 
some cases new bow shocks were revealed. One interesting example is 
the feature around HD 34078 (HIP 24575). The IRAS data, discussed 
in NC97, revealed an excess of emission at 60$\mu$m, but no 
discernible bow shock. With the WISE bands, two structures can be 
clearly seen (Fig. 2). There are filaments around the star mainly 
at longer wavelengths (red=22.2 microns), and a typical bow-shaped 
feature becomes visible in the longer bands (red=22.2 microns and 
green=12.1 microns) in the direction of the stellar motion. Another
example from group~1 is the object related to the B5 III star HD 
22928 (HIP 17358). The combination of low mass-loss rate and wind
terminal velocity, with a high stellar velocity leads to the formation
of a bow shock very close to the star. This situation results in a 
very large derived value of the ISM density (Table 4). The stellar 
motion is dominated by the tangential velocity, and is a nearby star. 
These help us to more clearly resolve the bow shock.

The strategy we adopted here for group~2, beginning with the early-type 
runaways, which are most likely to form bow shocks, has been successful
in identifying a significant number of new bow shocks. These results 
support the hypothesis that some runaway stars (around 10\%) produce 
detectable bow shocks. There are many possible reasons why a detectable
bow shock is not formed, such as a low-density ISM, an extremely high 
stellar velocity, or a low mass-loss rate, the inclination of the 
stellar velocity vector with respect to the plane of the sky, and 
confusion with strong field sources.

The current release of WISE data covers around 57 per cent of the sky,
but covers a slightly larger fraction of the Galactic plane. For our 
group 2 sample (244 stars), around 67 per cent of the stars were covered
in the first WISE data release (164 stars). A total of 17 of these were
found to have bow shocks (a detection rate of $\sim$ 10 per cent). 
On the basis of these statistics, we would expect to find around 8 
additional bow shock candidates in the remaining 80 group 2 sources.

Returning to the detected BS-C, several bow shocks in both group~1
and group~2 show a complex layered structure, such as HD 30614 (HIP
22783), HD 42933 (HIP 29276), and HD 15629 (HIP 11891). In many 
cases, but not all, the structures are aligned with the stellar 
velocity. An example of a misaligned bow shock candidate is HD 36512 
(HIP 25923).  We include this object in the sample because at this 
stage we lack any precise knowledge of the true stellar velocity.

HD 48099 (HIP 32076) is maybe the most classic example of a bow shock,
and this object will be the subject of a future detailed study that 
will include modeling of the infrared emission in the formation of a 
bow shock del Valle et al., in preparation).

With this sample, we will be able to investigate the IR luminosity 
and dust temperature and compare the scaling of these quantities 
with stellar parameters (van Buren \& McCray 1998), and also 
detectability (Stevens et al., in preparation). 

The current sample is insufficiently large to distinguish properties
according to stellar luminosity class, binarity status, or particular 
stellar classes, such as Of-type stars. However, when new data are 
released, the situation will improve.


\section{Summary and prospects}

We have discovered a significant number of bow shock candidates 
around early-type runaway stars providing a higher quality sample 
of bow shocks, the E-BOSS sample. For the set, we have determined 
a number of parameters of these objects. We have found no strong 
trends concerning the frequency of bow shocks with stellar mass, 
position, age, velocity, and spectral type.

In terms of future work, extending our study to later spectral types
will allow us to systematically search for and investigate radiative 
bow shocks (see for example G\'asp\'ar et al. 2008). The extensive 
Tetzlaff et al. (2010) database will also serve as a starting point 
for that study. We will also incorporate in forthcoming E-BOSS
versions other runaway databases together with individual objects 
from the literature.  

Detailed studies of individual objects will help us to more clearly
understand the stellar winds of the bow-shock producer stars, the 
medium in which they travel, and the stellar history, among other 
things. We plan to continue our studies of individual objects in 
the E-BOSS sample, as well as statistical studies of the global 
sample.

\begin{acknowledgements}
C.S. Peri and P.B. are supported by the ANPCyT PICT-2007/00848. P.B. 
also acknowledges support from CONICET PIP 0078 and UNLP G093 projects, 
and thanks the School of Physics and Astronomy of the University of 
Birmingham for kind hospitality. This publication makes uses of the 
NASA/IPAC Infrared Science Archive, which is operated by the Jet 
Propulsion Laboratory, California Institute of Technology, under 
contract with the National Aeronautics and Space Administration, 
the SIMBAD database, operated at CDS, Strasbourg, France, and data 
products from the Wide-field Infrared Survey Explorer, which is a 
joint project of the University of California, Los Angeles, and 
the Jet Propulsion Laboratory/California Institute of Technology, 
funded by the National Aeronautics and Space Administration. C.S. 
Peri is grateful to M.V. del Valle for a discussion on theoretical 
issues. We also thank an unknown A\&A referee for the comments and 
suggestions that have improved the article.

\end{acknowledgements}

\begin{table*}
\caption{Star members of Group 1.}
\label{ResultsNor-Cr}
\begin{minipage}{0.5\textwidth}
\centering
\begin{tabular}{l@{~~~}l@{~~~}l@{~~~}l@{~~~}l@{~~~}l}
\hline\hline
HIP & HD/BD/Other & Spectral type & MSX & WISE & 1997 \\
\hline
1415     &   1337                 & O9IIInn+... & --        & --         &             \\
2599     &   2905 / $\kappa$ Cas  & B1Iae       & $\star$   & $\supset$  & $\supset$   \\
3478     &   4142                 & B5V         & --        & $\star$    & $\bullet$   \\
13296    &  17505                 & O6Ve        & $\star$   & $\bullet$* & $\medcirc$? \\
14514    &  19374                 & B1.5V       & --        & $\star$    & $\bullet$   \\
15063    &  19820                 & O8.5III     & $\star$   & $\star$    & $\supset$?  \\
17358    &  22928                 & B5III       & --        & $\supset$  & $\supset$   \\
18370    &  24431                 & O9IV-V      & $\times$  & $\star$    & $\bullet$   \\
22783    &  30614 / $\alpha$ Cam  & O9.5Iae     & --        & $\supset$  & $\supset$   \\
24575    &  34078                 & O9.5Ve...   & --        & $\supset$  & $\bullet$   \\
25947    &  +39 1328              & O9III:      & $\times$  & $\star$    & $\bullet$   \\
-------- &  36862 / $\lambda$ Ori & B0.5V       & --        & [a]        & $\supset$   \\
26220    &  37020                 & B0.5V       & $\odot$   & $\odot$    & $\bullet$   \\
26889    &  37737                 & B0II:       & $\times$  & $\bullet$* & $\medcirc$? \\
28881    &  41161                 & O8V         & --        & $\supset$  & $\supset$   \\
29147    &  41997                 & O7.5V       & --        & $\odot$    & $\bullet$   \\
29276    &  42933 / $\delta$ Pic  & B3III+...   & --        & $\supset$  & $\supset$   \\
31978    &  47839                 & O7Ve        & $\star$   & $\odot$    &             \\
32067    &  48099                 & O6e         & $\star$   & $\supset$  & $\supset$   \\
33836    &  52533                 & O9V         & $\star$   & $\bullet$  & $\bullet$   \\
34536    &  54662                 & O7III       & $\star$   & $\supset$  & $\supset$   \\
35415    &  57061 / $\tau$ CMa    & O9Ib        & --        & $\odot$    & $\supset$   \\
38430    &  64315                 & O6e         & $\supset$ & $\odot$    & $\bullet$   \\
39429    &  66811                 & O4If(n)p    & --        & --         & $\bullet$   \\ 
50253    &  89137                 & O9.5III(n)p & $\times$  & --         & $\bullet$   \\
-------- &  92206                 & O6.5V       & [b]       & --         & $\bullet$   \\
56726    &  101131                & O6V((f))    & --        & $\odot$    & $\bullet$   \\
63117    &  112244                & O9Ibe       & --        & $\bullet$  & $\bullet$   \\
\hline
\end{tabular}
\end{minipage} \hfill
\begin{minipage}{0.5\textwidth}
\centering
\begin{tabular}{l@{~~~}l@{~~~}l@{~~~}l@{~~~}l@{~~~}l}
\hline\hline
HIP & HD/BD & Spectral type & MSX & WISE & 1997 \\
\hline
72510          &  130298                 & O5/O6       & $\star$    & $\supset$  & $\supset$  \\
74778          &  135240                 & O8.5V       & --         & $\bullet$  & $\bullet$  \\
77391          &  329905                 & O+...       & $\times$   & $\supset$  & $\bullet$  \\
78401          &  143275 / $\delta$ Sco  & B0.2IVe     & --         & $\supset$  & $\supset$  \\
81377          &  149757 / $\zeta$ Oph   & O9V         & --         & $\supset$  & $\supset$  \\
84588          &  156212                 & O+...       & $\star$    & $\star$    &            \\
85569$\dagger$ &  158186                 & O9.5V       & $\star$    & $\bullet$  & $\bullet$  \\
88333$\dagger$ &  164492                 & O6          & $\bullet$  & $\bullet$* & $\bullet$  \\
90320          &  169582                 & O5e         & $\times$   & $\bullet$  & $\bullet$  \\
91113          &  171491                 & B5          & $\star$    & $\odot$    & $\supset$  \\
92865          &  175514                 & O8:Vnn      & $\star$    & $\supset$  & $\supset$  \\
97280          &  186980                 & O7.5III...  & $\times$   & $\o$       & $\bullet$  \\
97796          &  188001                 & O7.5Ia...   & --         & $\supset$  & $\supset$  \\
98418          &  227018                 & O6.5III     & $\odot$    & --         & $\bullet$  \\
98530          &  189957                 & B0III       & --         & --         & $\bullet$  \\
101186         &  195592                 & O9.5Ia      & $\supset$  & --         & $\supset$  \\
---------      &  +43 3654 / U824        & O4I         & $\supset$  & --         & $\supset$  \\
103371         &  199579                 & O6V((f))    & $\star$    & --         & $\bullet$  \\
104642         &  202214                 & B0II        & --         & --         & $\bullet$  \\
105186         &  203064                 & O8e         & $\star$    & --         & $\bullet$  \\
105268         &  203467 / 6 Cep         & B3IVe       & --         & --         & $\supset$  \\
107598         &  207538                 & O9V         & $\bullet$  & --         & $\bullet$  \\
109556         &  210839 / $\lambda$ Cep & O6If(n)p    & $\star$    & --         & $\supset$  \\
110609         &  212593                 & B9Iab       & --         & --         & $\bullet$  \\
110817         &  213087                 & B0.5Ibe...  & --         & --         & $\bullet$  \\
111841         &  214680                 & O9V         & --         & --         & $\supset$? \\
114990         &  +63 1964               & B0II        & $\star$    & $\supset$  & $\bullet$  \\
117957         &  224151                 & B0.5II-III  & --         & $\star$    & $\bullet$  \\
\hline
\end{tabular}
\end{minipage}
\tablefoot{Stars of group 1 were taken from Noriega Crespo et al. (1997). 
All were observed by the IRAS satellite and two by the Spitzer-Glimpse 
program ($\dagger$). The first column lists the Hipparcos number of the 
star; the second one, other identification(s). Column (3) shows the
spectral classification, from the GOSC whenever possible or from NC97 
otherwise. Columns (4) and (5) give information about MSX and WISE 
emission on the stellar fields, according to the following symbols. 
$\star$: point source on star, 
$\supset$: bow shock candidate, 
$\supset$?: doubtful bow shock candidate, 
--: stellar field not covered by the survey, 
$\times$: no star or emission close-by, 
$\odot$: extended  source on star, 
$\bullet$*: confusion with larger structure, 
$\o$: diffuse emission,
$\bullet$: emission excess,
$\medcirc$?: possible bubble.
Column (6) lists the qualifiers (same meaning as before) as in the original
work by NC97. [a] The star HD 36862, together with HD 36861 (O8III), belong 
to a rich cluster, are not single, and have a large spatial velocity. More 
than one feature can produce the surrounding IR emission detected. [b] Cappa 
et al. (2008) studied the environs of HD 92206, and identified an HII region. 
We discarded these two cases from our sample.}
\end{table*}

\begin{table*}
\caption{Star members of group 2.}
\label{grupo2}
\begin{center}
\begin{tabular}{ll | ll | ll | ll | ll | ll}
\hline\hline
HIP & Sp.t. & HIP & Sp.t. & HIP & Sp.t. & HIP & Sp.t. & HIP & Sp.t. & HIP & Sp.t.\\
\hline
&&&&&&&&&&&\\
278         & B2IV     & 505          & O6pe     & 1805        & B0IV     & {\bf 2036}  & B1V    & {\bf 2599}  & B1Ia     & 4532        & B1II    \\
4983        & B2IV-V   & 5391         & B1V      & 6027        & B2III    & 8725        & O8V    & 9538        & B1V      & 10463       & B2IV-V  \\
10527       & B0.5III  & 10641        & B2Ib     & 10849       & B2V      & 10974       & B2     & 11099       & O8.5V    & 11279       & B2Ia    \\
11347       & B1Ib     & 11394        & O6       & 11396       & B2       & 11473       & O9.5V  & 11792       & O9V      & {\bf 11891} & O5      \\
12009       & B1Iab    & 12293        & B2       & 13736       & B0II-III & 13924       & O7V    & 14514       & B1.5V    & 14626       & B1V     \\
14777       & B2       & 14969        & B2IV     & 15270       & B2.5IV-V & {\bf 16518} & B1V    & 16566       & O        & 17387       & B2V     \\
18151       & B1III    & 18350        & O9.5     & 18614       & O7.5Iab  & 19218       & O8     & 21626       & B2.5V    & 22061       & B2.5V   \\
22461       & B1II-III & 23060        & B2V      & 24072       & B2III    & 24238       & B2V    & {\bf 24575} & O9.5V    & {\bf 25923} & B0V     \\
26064       & B2IV-V   & {\bf 26397}  & B0.5V    & 26889       & B0II     & 27204       & B1IV-V & 27850       & B1V      & 27941       & O6      \\
28756       & B2V      & 29201        & B0V      & {\bf 29276} & B0.5IV   & 29317       & B1V    & 29321       & B2V      & 29563       & B2V     \\
29678       & B1V      & 30961        & B2.5IV-V & {\bf 31766} & O9.5II   & 31787       & B0IV   & {\bf 32067} & O6       & 32300       & B0.5IV  \\
32602       & O6       & 32947        & B2V      & 33300       & B2V      & 33754       & B1Ib   & {\bf 34536} & O6       & 34924       & B2III   \\
34986       & B0.5III  & 35149        & B1.5III  & 35951       & B2V      & 36369       & O6     & 36778       & B2V      & 37169       & O9.5Iab \\
38855       & B2V      & 39172        & B2.5V    & 40047       & O5p      & 44685       & B2IV   & 45880       & B2       & 46760       & B2V     \\
48715       & B1Ib     & 52670        & B2.5V    & 54572       & B2V      & 58748       & B1II   & 61431       & B1Ib     & {\bf 62322} & B2V     \\
62829       & B0.5III  & 63049        & B0IV     & 63117       & O9Ib     & 63170       & B0.5Ia & 63256       & B2V      & 64272       & B1Ib    \\
67663       & B2V      & 68002        & B2.5IV-V & 68817       & B0.5V    & 69892       & O8.5   & 69996       & B2.5IV   & 70574       & B2IV    \\
70877       & B2III    & 71264        & B2V      & 72438       & B2.5V    & {\bf 72510} & O7.5   & 72710       & B2       & 74778       & O8.5V   \\
{\bf 75095} & B2Ib     & 75141        & B1.5IV   & 75711       & B2II/III & 76013       & B1     & 76642       & B2III    & 78145       & B0.5Ia  \\
78582       & B2V      & 79466        & B2III    & 80782       & B1.5Iap  & 80945       & B1Ia   & 81100       & O6e      & 81122       & B0Ia    \\
81305       & O9Ia     & {\bf 81377}  & O9.5V    & 81696       & O7V      & {\bf 82171} & B0Iab  & 82378       & O9.5IV   & 82691       & O7e     \\
82775       & O8Iab... & 82783        & O9Ia     & 83003       & O...     & 83574       & B2Iab  & 83635       & B1V      & 84226       & B1Ib    \\
84338       & B2III    & 84401        & O9       & 84687       & B0V      & 84745       & B2V    & 85331       & O6.5III  & 85530       & B2V     \\
85885       & B2II     & 87397        & B2III    & 88004       & B1Iab    & 88496       & B2V    & 88584       & O6       & {\bf 88652} & O9.5Iab \\
88714       & B2Ib     & 89743        & O9.5V    & 90610       & B2V      & 90804       & B2V    & 90950       & B0Ia/Iab & 91003       & O7      \\
91049       & B2II     & 91599        & B0.5V    & 92133       & B2.4V    & 93118       & O7.5   & 93796       & B1Ib     & 93934       & B2II    \\
94934       & B2IV     & 95408        & B2V      & 96130       & B1.5III  & 96362       & B2V    & 97246       & B1Ia     & 97545       & B1V     \\
97679       & B2.5V    & 117514       & B2V      &             &          &             &        &             &          &             &         \\
&&&&&&&&&&&\\
\hline
&&&&&&&&&&&\\
3013   & B2      & 38518  & B0.5Ib   & 39429  & O8Iaf   & 39776  & B2.5III & 40341  & B2V      & 41168  & B2IV   \\
41463  & B2V     & 41878  & B1.5Ib   & 42316  & B1Ib    & 42354  & B2III   & 43158  & B0II/III & 43868  & B1Ib   \\
44251  & B2.5V   & 44368  & B0.5Ib   & 46950  & B1.5IV  & 47868  & B0IV    & 48469  & B1V      & 48527  & B2V    \\
48730  & B2IV-V  & 48745  & B2III    & 49608  & B1III   & 49934  & B2IV    & 50899  & B0Iab/Ib & 51624  & B1Ib   \\
52526  & B0Ib    & 52849  & O9V      & 52898  & B2III   & 54179  & B1Iab   & 54475  & O9II     & 58587  & B2IV   \\
61958  & Op      & 65388  & B2       & 74368  & B0      & 89902  & B2V     & 94716  & B1II-III & 97045  & B0V    \\
97845  & B0.5III & 98418  & O7       & 98661  & B1Iab   & 99283  & B0.5IV  & 99303  & B2.5V    & 99435  & B0.5V  \\
99580  & O5e     & 99953  & B1V      & 100088 & B1.5V   & 100142 & B2V     & 100314 & B1.5Ia   & 100409 & B1Ib   \\
101186 & O9.5Ia  & 101350 & B0V      & 102999 & B0IV    & 103763 & B2V     & 104316 & O9       & 104548 & B1V    \\
104579 & B1V     & 104814 & B0.5V    & 105186 & O8      & 105912 & B2II    & 106620 & B2V      & 106716 & B2V    \\
107864 & Op      & 108911 & B2Iab    & 109051 & B2.5III & 109082 & B2V     & 109311 & B1V      & 109332 & B2III  \\
109556 & B1II    & 109562 & O9Ib     & 109996 & B1II    & 110025 & B2III   & 110287 & B1V      & 110362 & B0.5IV \\
110386 & B2IV-V  & 110662 & B1.5IV-V & 110817 & B0.5Ib  & 111071 & B0IV    & 112482 & B1II     & 112698 & B1V    \\
114482 & O9.5Iab & 114685 & O7       &        &         &        &         &        &          &        &        \\
&&&&&&&&&&&\\
\hline
\end{tabular}
\tablefoot{Top: Stars {\sl with} WISE observations (164 stars). Bottom: stars {\sl without} WISE observations (80 stars).
The spectral types are from Tetzlaff et al. (2010). In bold font: the 17 bow shock candidates detected, see text.}
\end{center}
\end{table*}

\begin{table*}
\caption{List of the E-BOSS bow shock candidates and corresponding stellar parameters.}
\centering
\begin{tabular}{l@{~~~}l@{~~~}c@{~~~}c@{~~~}l@{~~~}r@{~~~}r@{~~~}r@{~~~}r@{~~~}r@{~~~}r@{~~~}r}
\hline\hline
Star & Group  & $l$     & $b$      & Spectral type & $d$  & $v_{\infty}$  & $\dot{M} \times 10^{6}$   & $v_{\rm tg}$  & $v_{\rm r}$   & $\mu_{\alpha} \cos \delta$ & $\mu_{\delta}$ \\
     &        & [\degr] & [\degr]  &               & [pc] & [km s$^{-1}$] & [$M_{\odot}$ yr$^{-1}$] & [km s$^{-1}$] & [km s$^{-1}$] & [mas yr$^{-1}$]            & [mas yr$^{-1}$] \\
\hline
HIP 2036   & 2   & 120.9137 & +09.0357 & O9.5III+B1V    & 757$\pm$161$^{\rm a}$  & [1200]  & 0.48     &  15.2   & $-$5     & $-$1.66  & 1.90     \\
HIP 2599   & 1,2 & 120.8361 & +00.1351 & B1 Iae         & 1457$\pm$300$^{\rm a}$ & 1105    & 0.12     &  26.2   & $-$2.3   &    3.65  & $-$2.07  \\
HIP 11891  & 2   & 134.7692 & +01.0144 & O5 V((f))      & ( 900 )                & 2810    & 1.10     &  11.9   & $-$48    &    0.03  & $-$2.16  \\
HIP 16518  & 2   & 156.3159 & -16.7535 & B1 V           & ( 650 )                & [500]   & 0.006    &  47.3   & 25       & $-$8.28  & 3.44     \\
HIP 17358  & 1   & 150.2834 & -05.7684 & B5 III         & ( 150 )                & [500]   & $<$0.001 &  [35]   & 4        &   25.58  & $-$43.06 \\
HIP 22783  & 1   & 144.0656 & +14.0424 & O9.5 Ia        & 1607$\pm$275$^{\rm a}$ & 1590    & 0.25     &  [52]   & 6.1      & $-$0.13  & 6.89     \\
HIP 24575  & 2   & 172.0813 & -02.2592 & O9.5 V         & 548$\pm$68$^{\rm a}$   & [1200]  & 0.1      & 140.0   & 59.1     & $-$3.58  & 43.73    \\
HIP 25923  & 2   & 210.4356 & -20.9830 & B0 V           & ( 900 )                & [1000]  & 0.06     &  16.8   & 17.4     & $-$0.10  & $-$4.87  \\
HIP 26397  & 2   & 174.0618 & +01.5808 & B0.5 V         & ( 350 )                & [750]   & 0.014    &  11.9   & $-$19    &  0.88    & $-$3.61  \\
HIP 28881  & 1   & 164.9727 & +12.8935 & O8 Vn          &  1500$^{\rm b}$        & 2070    & 0.03     &  [17]   &  5       & $-$0.82  & $-$1.49  \\
HIP 29276  & 1,2 & 263.3029 & -27.6837 & B1/2 III       & ( 400 )                & [600]   & $<$0.001 &   9.2   & 30.6     & $-$4.90  & 7.41     \\
HIP 31766  & 2   & 210.0349 & -02.1105 & O9.7 Ib        & 1414$\pm$28$^{\rm a}$  & 1590    & 1.07     &   6.7   & 58.4     & $-$0.34  & $-$0.83  \\  
HIP 32067  & 1,2 & 206.2096 & +00.7982 & O5.5V((f))+... & 2117$\pm$367$^{\rm a}$ & 2960    & 0.13     &  23.4   & 31       &    0.84  & 2.55     \\
HIP 34536  & 1,2 & 224.1685 & -00.7784 & O6.5V((f))+... & 1293$\pm$206$^{\rm a}$ & 2456    & 0.19     &  14.3   & 58       & $-$1.96  & 4.40     \\
HIP 38430  & 1   & 243.1553 & +00.3630 & O6Vn+...       & ( 900 )                & [2570]  & 0.7      &  [13]   & 28       & $-$3.04  & $-$0.38  \\
HIP 62322  & 2   & 302.4492 & -05.2412 & B2.5 V         & ( 150 )                & [300]   & 0.006    &   4.5   & 42       & $-$41.97 & $-$8.89  \\
HIP 72510  & 1,2 & 318.7681 & +02.7685 & O6.5III(n)(f)  & ( 350 )                & [2545]  & 0.27     &   7.4   & $-$74    & $-$7.49  & $-$5.15  \\
HIP 75095  & 2   & 322.6802 & +00.9060 & B1Iab/Ib       & ( 800 )                & [1065]  & 0.14     &  28.6   & 4        & $-$8.42  & $-$9.18  \\
HIP 77391  & 1   & 330.4212 & +04.5928 & O9 I           & ( 800 )                & [1990]  & 0.25     &  [19]   & 15       & $-$4.63  & $-$1.84  \\
HIP 78401  & 1   & 350.0969 & +22.4904 & B0.2 IVe       & 224$\pm$24$^{\rm a}$   & [1100]  & 0.14     &  [38]   & $-$7     & -10.21   & -35.41   \\
HIP 81377  & 1,2 & 006.2812 & +23.5877 & O9.5 Vnn       & 222$\pm$22$^{\rm a}$   & [1500]  & 0.02     &  24.4   & $-$15    & 15.26    & 24.79    \\
HIP 82171  & 2   & 329.9790 & -08.4736 & B0.5 Ia        & 845$\pm$120$^{\rm a}$  & 1345    & 0.09     &  65.7   & $-$53.3  & $-$4.64  & $-$20.28 \\
HIP 88652  & 2   & 015.1187 & +03.3349 & B0 Ia          & ( 650 )                & [1535]  & 0.5      &   8.2   & 30       & $-$1.05  & $-$1.38  \\
HIP 92865  & 1   & 041.7070 & +03.3784 & O8 Vnn         & ( 350 )                & [1755]  & 0.04     &   [2]   & $-$41    & $-$0.78  & 0.46     \\
HIP 97796  & 1   & 056.4824 & -04.3314 & O7.5 Iabf      & 2200$^{\rm c}$         & [1980]  & 0.50     & [110]   & 9        &  $-$2.03 & $-$10.30 \\
HIP 101186 & 1   & 082.3557 & +02.9571 & O9.7 Ia        & 1486$\pm$402$^{\rm a}$ & [1735]  & 0.23     &  22.3   & $-$28    & $-$2.37  & 1.37     \\
BD+43 3654 & 1   & 082.4100 & +02.3254 & O4 If          & 1450$^{\rm d}$         & [2325]  & 6.5      &  [14]   & $-$66.2  & $-$0.44  & 1.3      \\
HIP 114990 & 1   & 112.8862 & +03.0998 & B0 II          & 1400$^{\rm e}$         & [1400]  & 0.6      &  [52]   & $-$125.3 & $-$7.86  & $-$0.71  \\
\hline
\end{tabular}
\tablefoot{Galactic coordinates: taken from Simbad. Spectral types: for B-type stars from the Simbad database, 
for O-type stars GOS Catalog. References for the distance values: (a) Megier et al. (2009), (b) Mason et al. 
(1998), (c) Schilbach \& Roeser (2008), (d) Hanson (2003), (e) Thorburn et al. (2003); distances in 
brackets: derived from Hipparcos (van Leeuwen 2007) parallaxes. Terminal velocities in square brackets: from 
Howarth et al. (1997), otherwise inter- or extrapolated from Prinja et al. (1990). Mass-loss rates: derived 
from Vink et al. (2001). Tangential velocities in brackets derived from proper motions (van Leeuwen 2007), 
otherwise from Tetzlaff et al. (2010). Radial velocities are from the Second Catalog of Radial Velocities 
with Astrometric Data (Kharchenko et al. 2007).}
\end{table*}

\begin{table*}
\caption{Observational parameters of the bow shock candidates obtained from the images of MSX and WISE.} 
\centering
\begin{tabular}{l@{~~~}|l@{~~~}l@{~~~}l@{~~~}|l@{~~~}l@{~~~}l@{~~~}|r@{~~~}|l}
\hline\hline
Star       & $l$ & $w$  & $R$ &  $l$ & $w$ & $R$ & $n_{\rm ISM}$  & Comments \\
           &   & [arcmin] &   &  & [pc]  &       & [cm$^{-3}$]    &         \\
\hline
HIP 2036   & 4.5 & 1.3	& 1	& 0.99	& 0.29	& 0.22	&  130  & Emission-line star      \\
HIP 2599   & 9	 & 1.3	& 3	& 3.81	& 0.55	& 1.27	&  0.4  & Emission-line star      \\
HIP 11891  & 4	 & 1	& 1	& 1.05	& 0.26	& 0.26	&    3  & Star in cluster         \\
HIP 16518  & 4	 & 1	& 0.7	& 0.76	& 0.19	& 0.13	&  0.2  & Variable star           \\
HIP 17358  & 3	 & 1	& 1	& 0.13	& 0.04	& 0.04	&  600  & Variable star           \\
HIP 22783  & 33	 & 10	& 10	& 15.43	& 4.67	& 4.67	&  0.02 & Emission-line star      \\
HIP 24575  & 2	 & 0.5	& 0.4	& 0.32	& 0.08	& 0.06	&    3  & Double or multiple star \\
HIP 25923  & 4	 & 1	& 1.5	& 1.05	& 0.26	& 0.39	&    1  & Variable star           \\
HIP 26397  & 3	 & 1	& 1	& 0.31	& 0.10	& 0.10	&    2  & Star                    \\
HIP 28881  & 9	 & 1.5	& 3	& 5.54	& 0.92	& 1.85	&  0.3  & Double or multiple star \\
HIP 29276  & 5	 & 2	& 2	& 0.58	& 0.23	& 0.23	& 0.003 & Eclipsing binary of beta Lyr type \\
HIP 31766  & 5	 & 2	& 2	& 2.06	& 0.82	& 0.82	&  0.03 & Double or multiple star \\
HIP 32067  & 13	 & 2.5	& 3	& 8.01	& 1.54	& 1.85	&  0.1  & Emission-line star      \\
HIP 34536  & 12	 & 3	& 4	& 4.51	& 1.13	& 1.50	&  0.01 & HII (ionized) region    \\
HIP 38430  & 2	 & 0.5	& 0.5	& 0.52	& 0.13	& 0.13	&   60  & Emission-line star      \\
HIP 62322  & 4	 & 1.2	& 1	& 0.17	& 0.05	& 0.04	&  0.02 & Double or multiple star \\
HIP 72510  & 4.5 & 0.8	& 1.5	& 0.46	& 0.08	& 0.15	&  0.2  & Emission-line star      \\
HIP 75095  & 1.5 & 0.5	& 0.5	& 0.35	& 0.12	& 0.12	&   40  & Star                    \\
HIP 77391  & 4	 & 1	& 1	& 0.93	& 0.23	& 0.23	&   30  & Star                    \\
HIP 78401  & 25	 & 2	& 6	& 1.63	& 0.13	& 0.39	&    2  & Double or multiple star \\
HIP 81377  & 22	 & 2	& 5	& 1.42	& 0.13	& 0.32	&    1  & Be star                 \\
HIP 82171  & 2	 & 0.5	& 0.7	& 0.49	& 0.12	& 0.17	&    1  & Star                    \\
HIP 88652  & 6	 & 1	& 1.5	& 1.13	& 0.19	& 0.28	&    2  & Star                    \\
HIP 92865  & 11	 & 1	& 3     & 1.12	& 0.10	& 0.31	&  0.003& Eclipsing binary of beta Lyr type \\
HIP 97796  & 13	 & 2.5	& 6	& 8.32	& 1.60	& 3.84	&  0.02 & Spectroscopic binary    \\
HIP 101186 & 19	 & 2.5	& 4	& 8.21	& 1.08	& 1.73	&  0.1  & Emission-line star      \\
BD+43 3654 & 12	 & 3	& 3.5	& 5.06	& 1.27	& 1.48	&  0.2  & Star                    \\
HIP 114990 & 3.5 & 0.75	& 1.5	& 1.43	& 0.31	& 0.61	&  0.05 & Star                    \\
\hline
\end{tabular}
\tablefoot{Columns (2) to (7): length ($l$) and width ($w$) of the bow shock structure, 
and distance ($R$) from the star to the midpoint of the bow shock, in angular and linear 
units. Column (8): the ambient density $n_{\rm ISM}$ (see text). The descriptions in the 
last column are taken from the Simbad database.}
\end{table*}

\include{bowshock-figures-3-PDF}

\include{estadistica-v2}

\end{document}

%% file: bowshock-figures-3-PDF.tex
\begin{figure*}[t]

\begin{minipage}{\textwidth}
\centering
\includegraphics[width=0.49\textwidth,height=7.3cm]{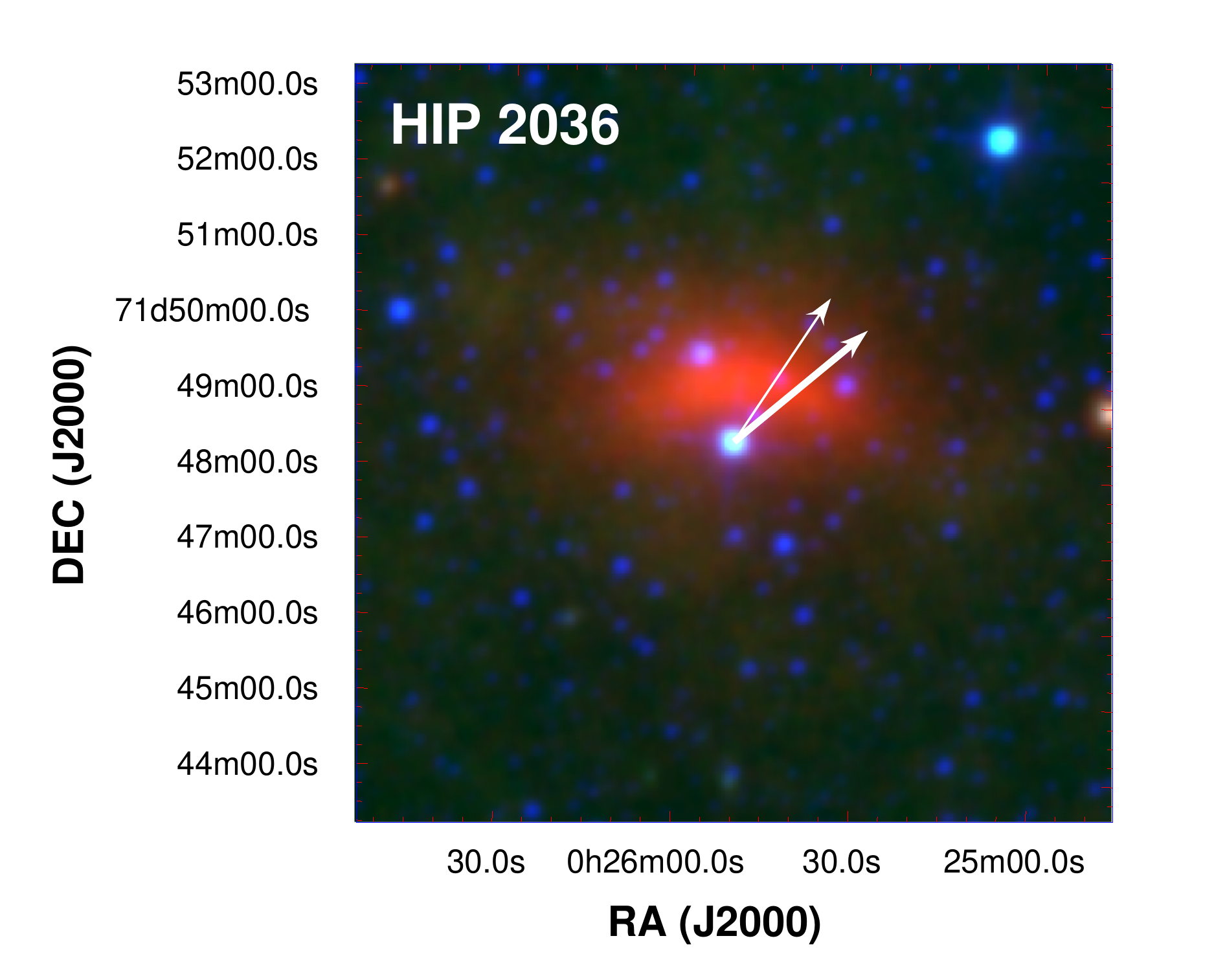}
\hfill
\includegraphics[width=0.45\textwidth,height=6.85cm]{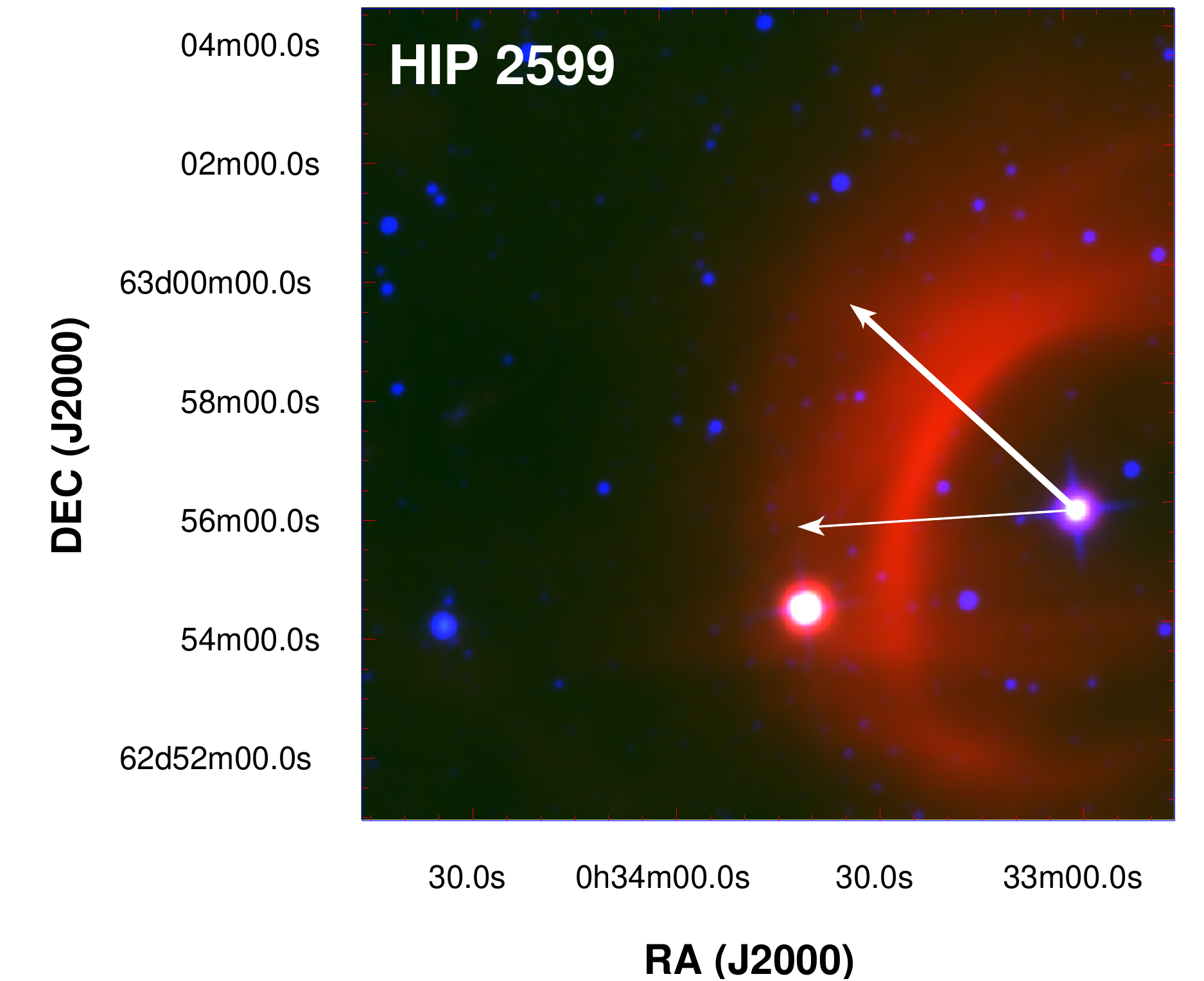}
\end{minipage} 

\begin{minipage}{\textwidth}
\centering
\includegraphics[width=0.49\textwidth,height=7.2cm]{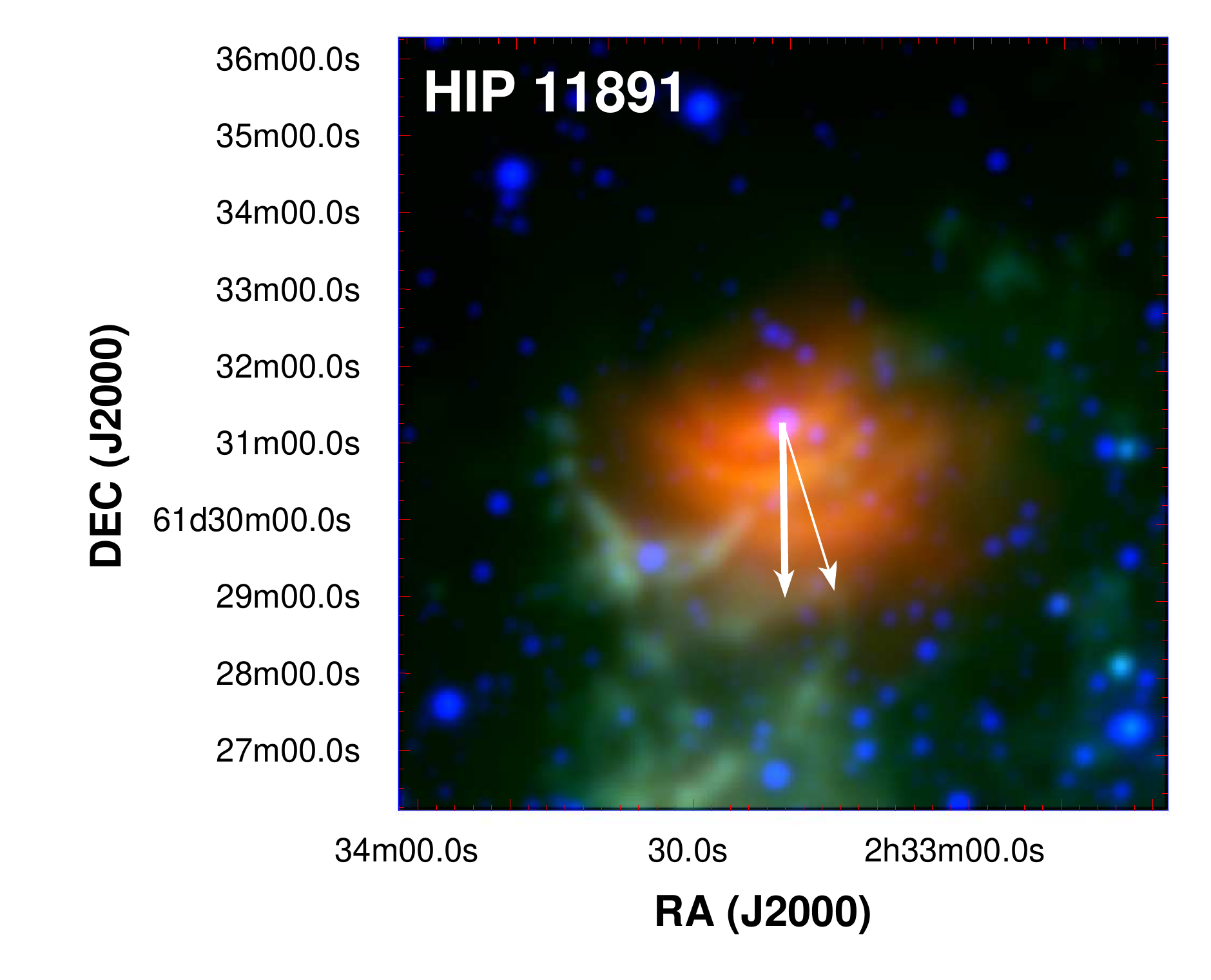}
\hfill
\includegraphics[width=0.46\textwidth,height=7.0cm]{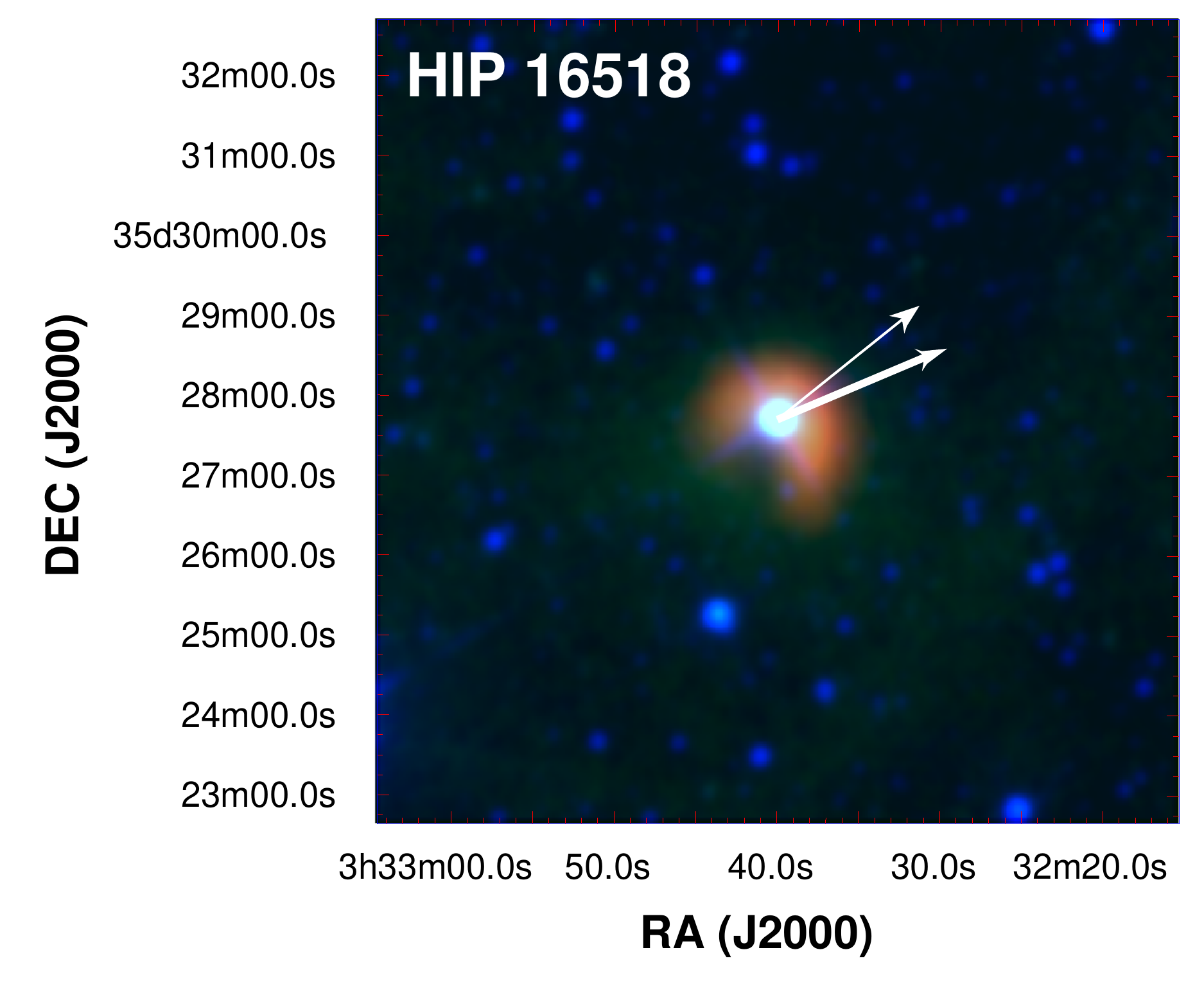}
\end{minipage} 

\begin{minipage}{\textwidth}
\centering
\includegraphics[width=0.44\textwidth,height=6.9cm]{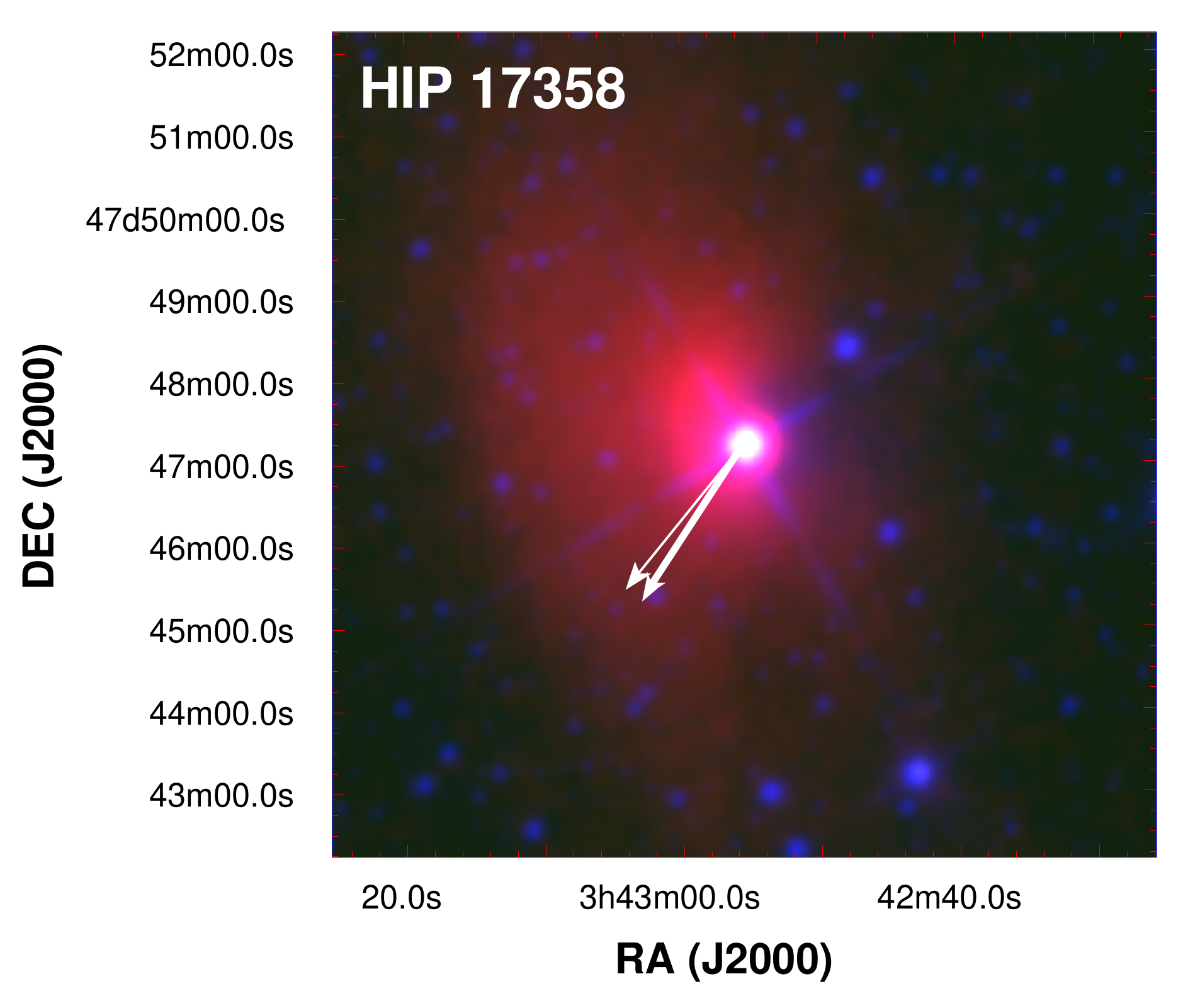}
\hfill
\includegraphics[width=0.52\textwidth,height=7.05cm]{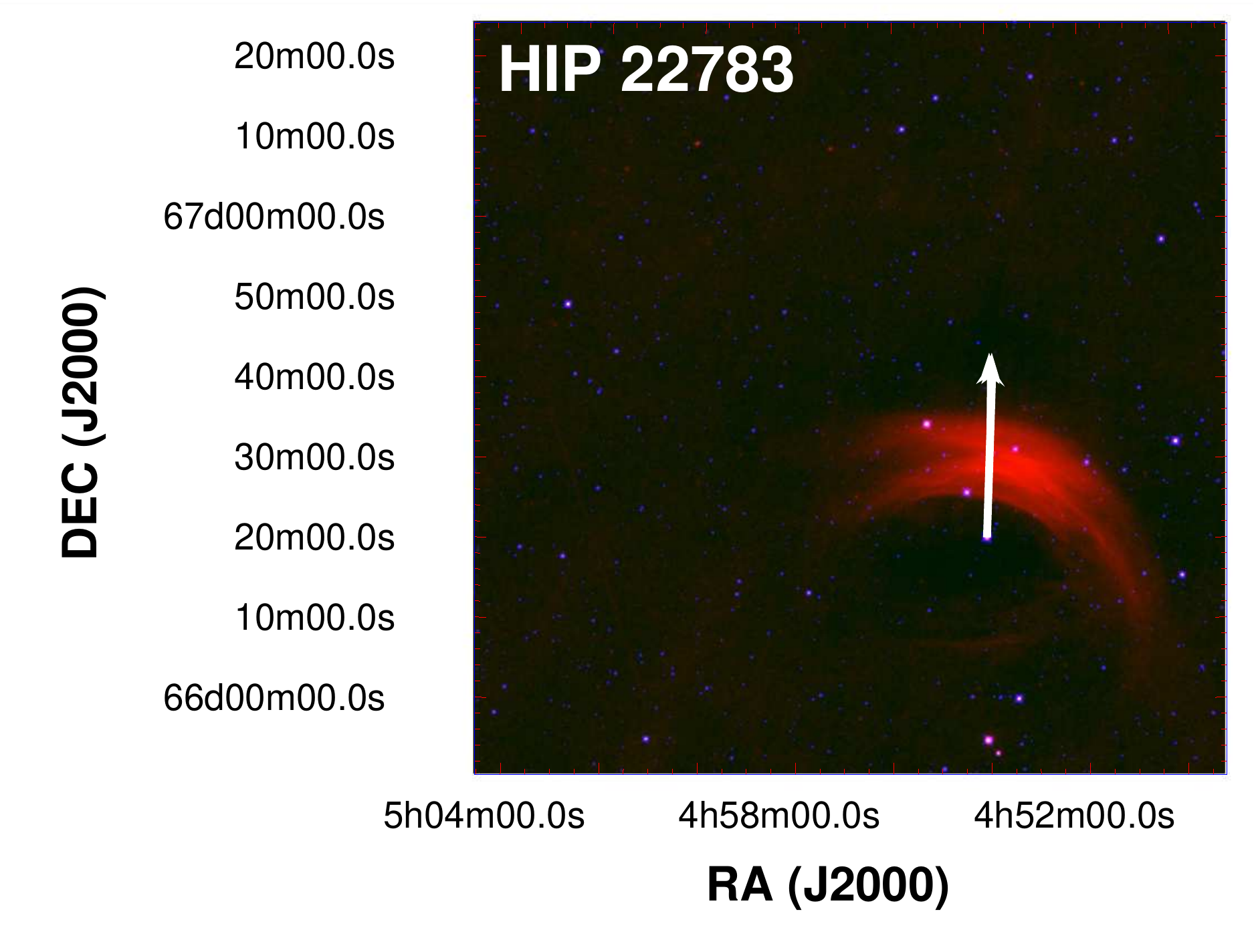}
\end{minipage}

\vspace{0.5cm}
\caption{WISE images of the bow shock candidates around 6 OB stars. 
Color mapping: {\bf blue}=3.4 microns; {\bf green}=12.1 microns;
{\bf red}=22.2 microns. The color scales are in data numbers (DNs);
the WISE data is not calibrated in their surface brightness.
The vectors indicate the direction of the star proper motion: the thicker one
represents the one derived from Hipparcos data by van Leeuwen 2007; 
the thinner one is the same but corrected for the ISM motion caused by Galactic rotation. The vectors length are not scaled with the original values.}
\label{bwshcks1to6}
\end{figure*}

\begin{figure*}[t]
\begin{minipage}{\textwidth}
\centering
\includegraphics[width=0.44\textwidth,height=6.8cm]{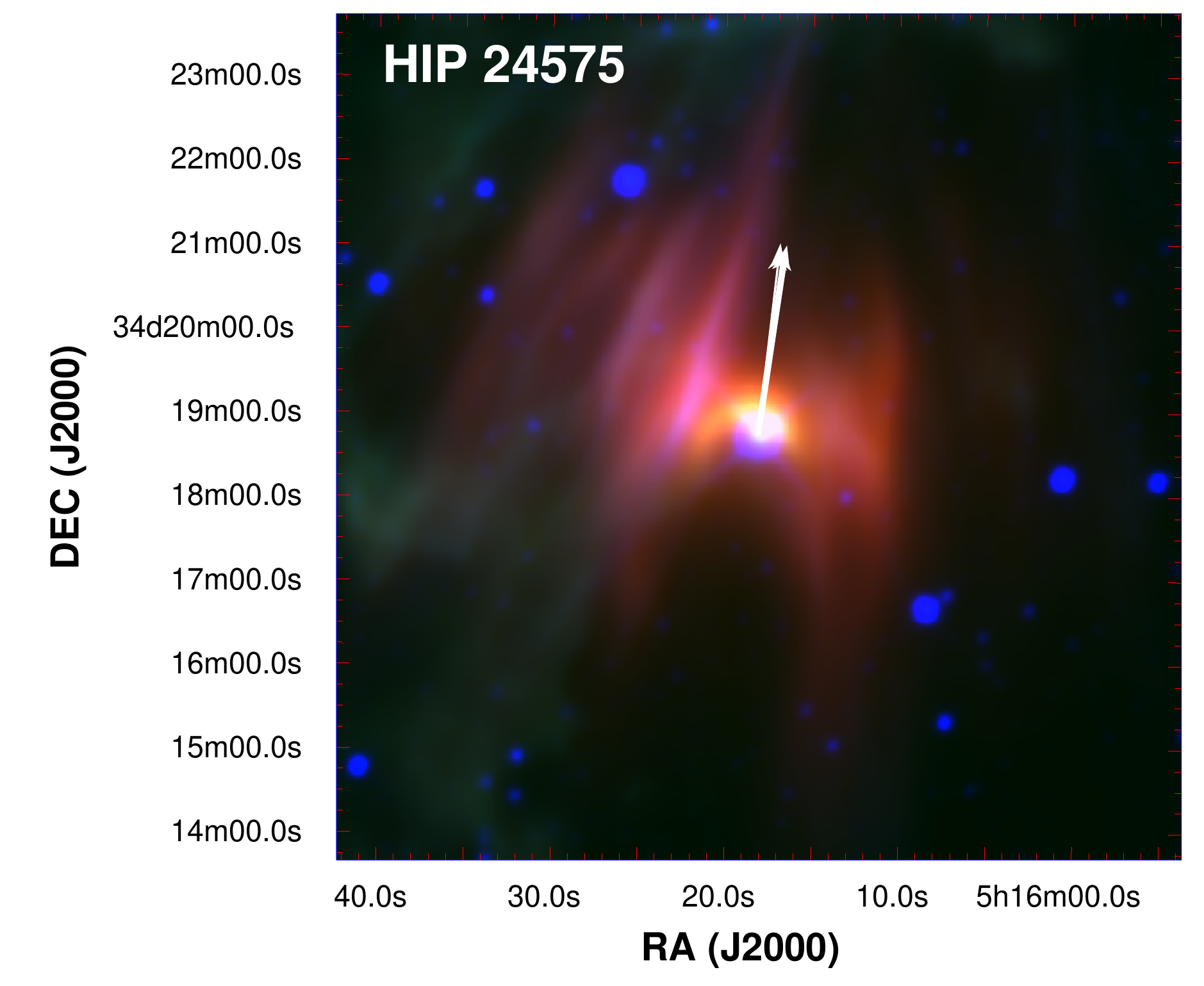}
\hfill
\includegraphics[width=0.45\textwidth,height=6.8cm]{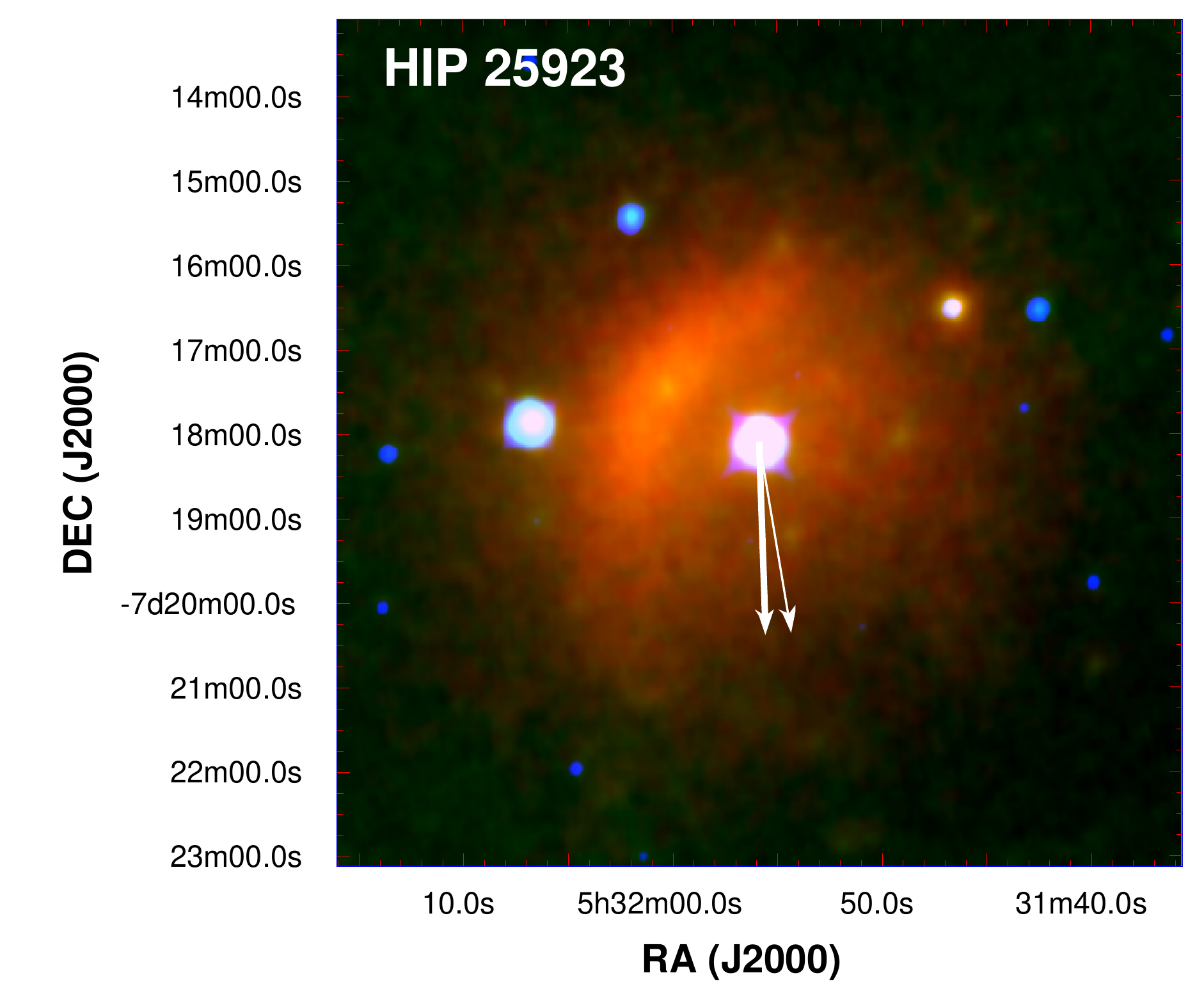}
\end{minipage} 

\begin{minipage}{\textwidth}
\centering
\includegraphics[width=0.46\textwidth,height=6.8cm]{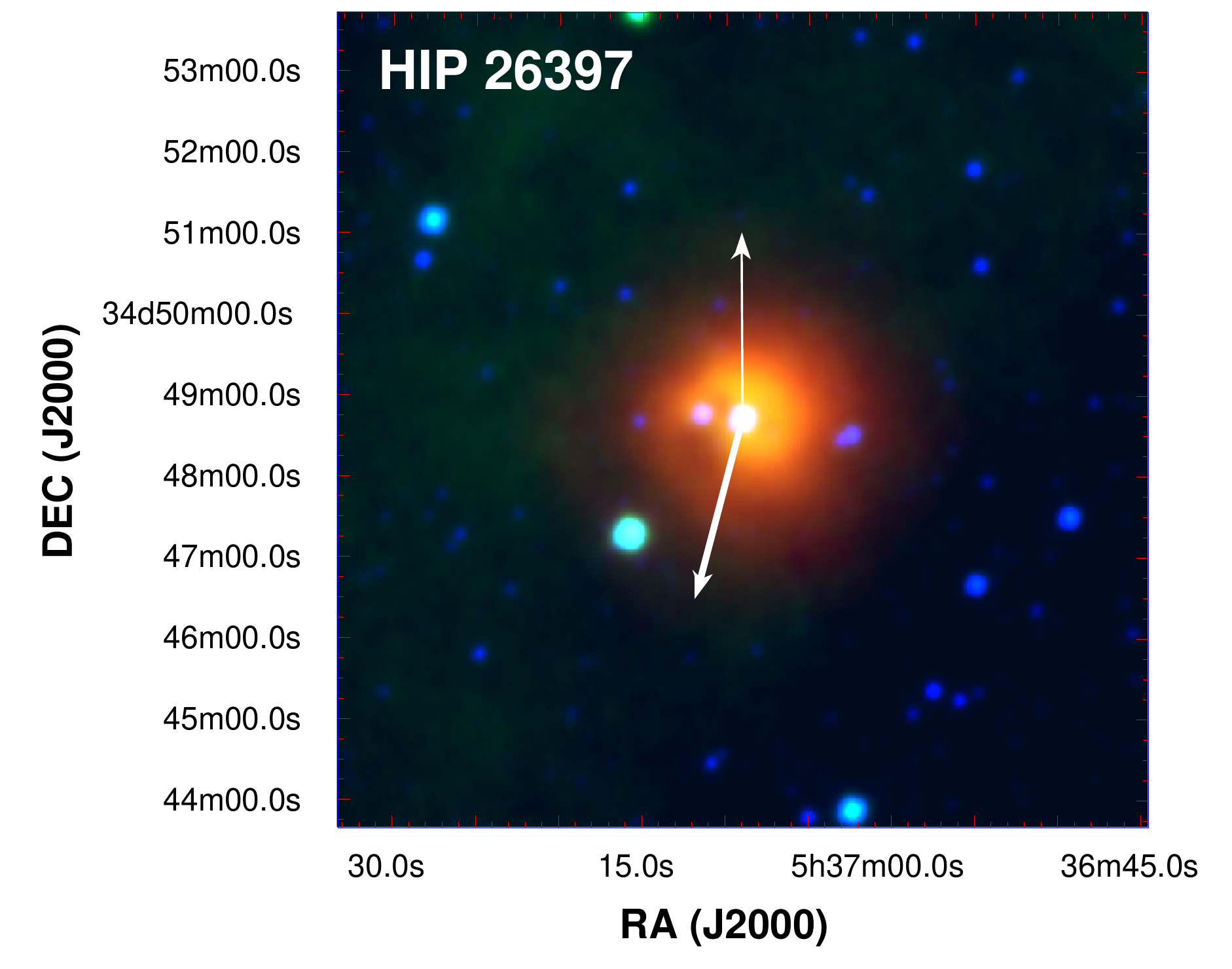}
\hfill
\includegraphics[width=0.46\textwidth,height=6.8cm]{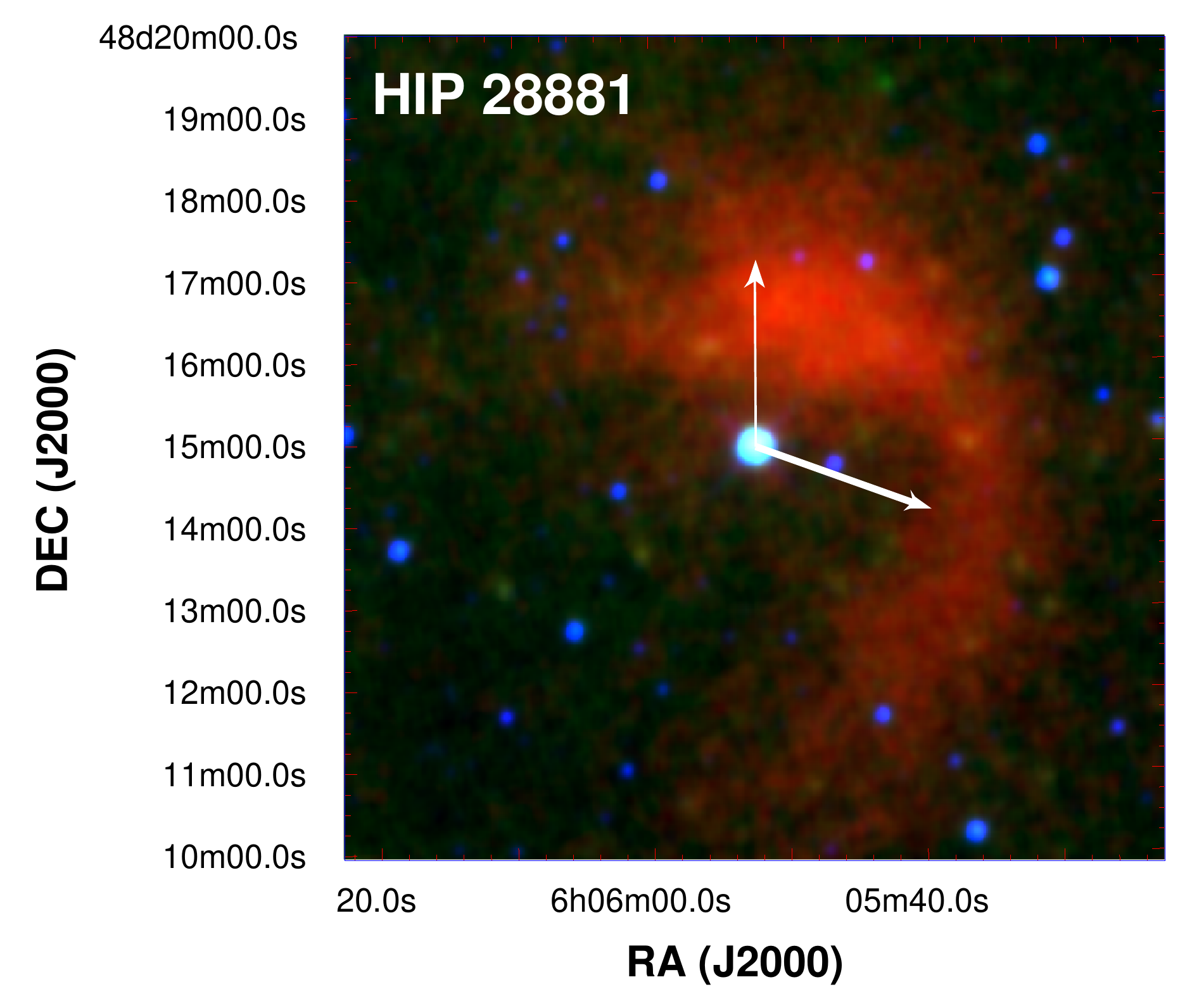}
\end{minipage} 

\begin{minipage}{\textwidth}
\centering
\includegraphics[width=0.45\textwidth,height=6.8cm]{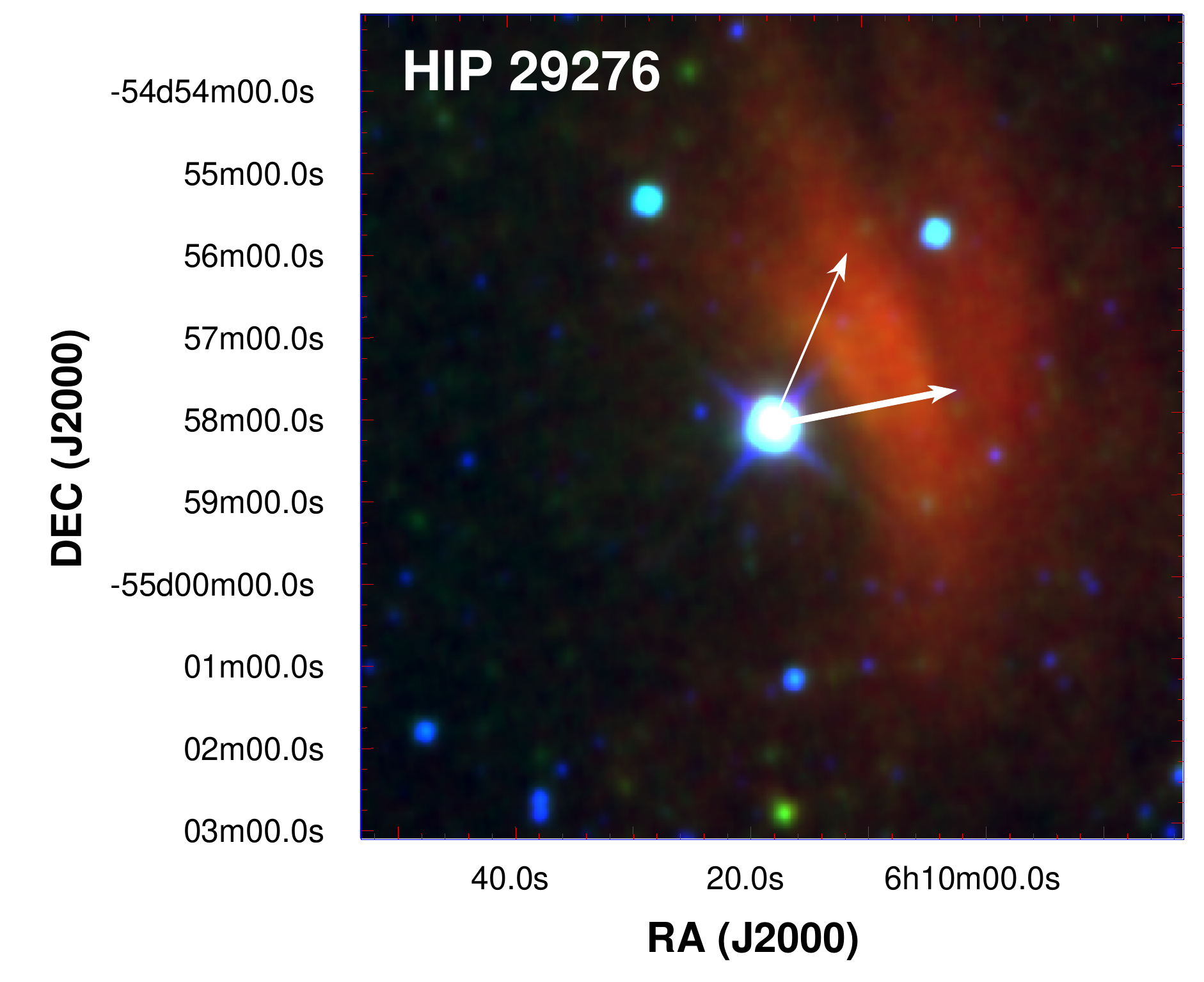}
\hfill
\includegraphics[width=0.45\textwidth,height=6.8cm]{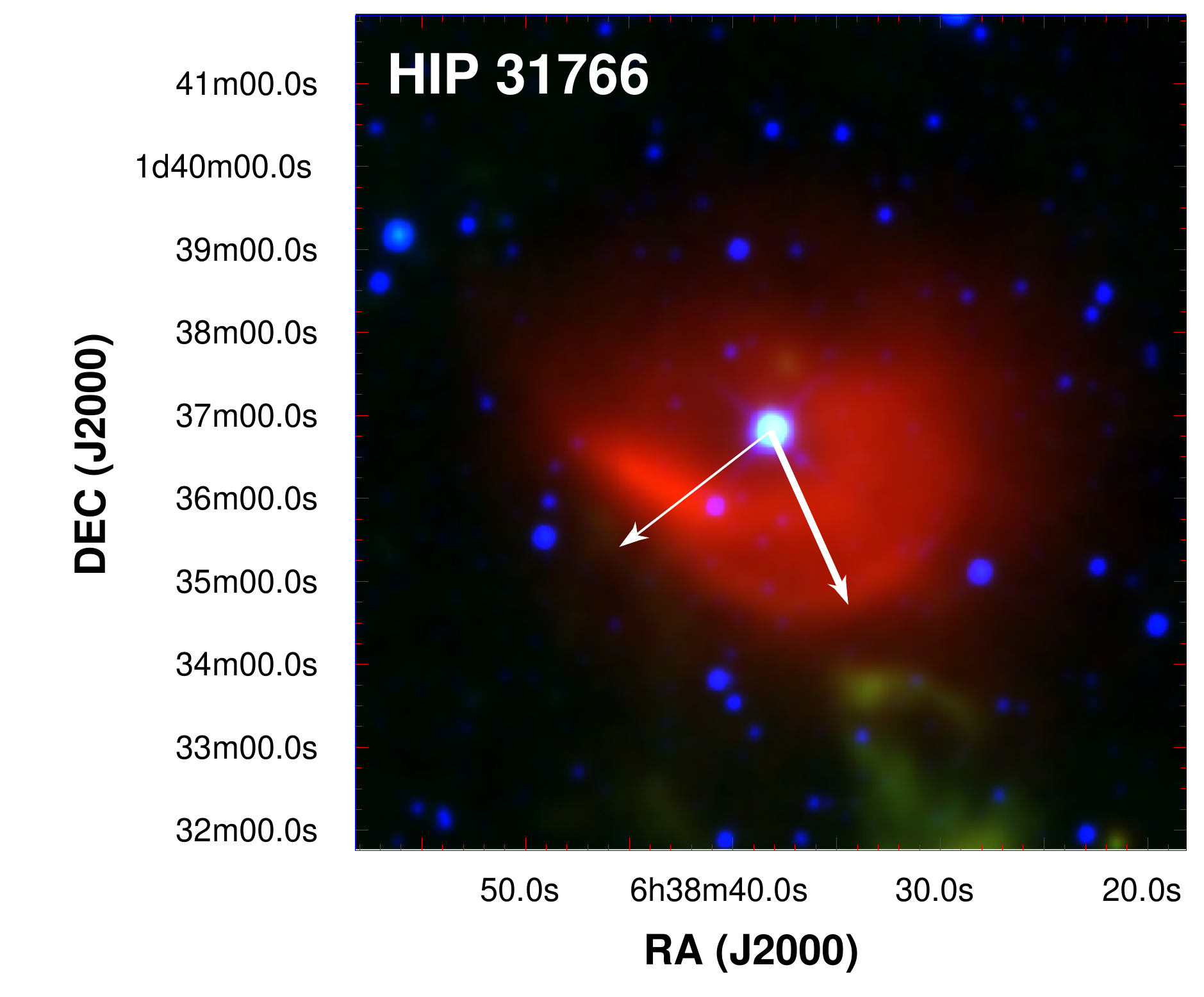}
\end{minipage} 

\vspace{0.5cm}
\caption{Same as in Figure 1, for another six OB stars.}
\label{bwshcks7to12}
\end{figure*}

\begin{figure*}[t]
\begin{minipage}{\textwidth}
\centering
\includegraphics[width=0.46\textwidth,height=6.5cm]{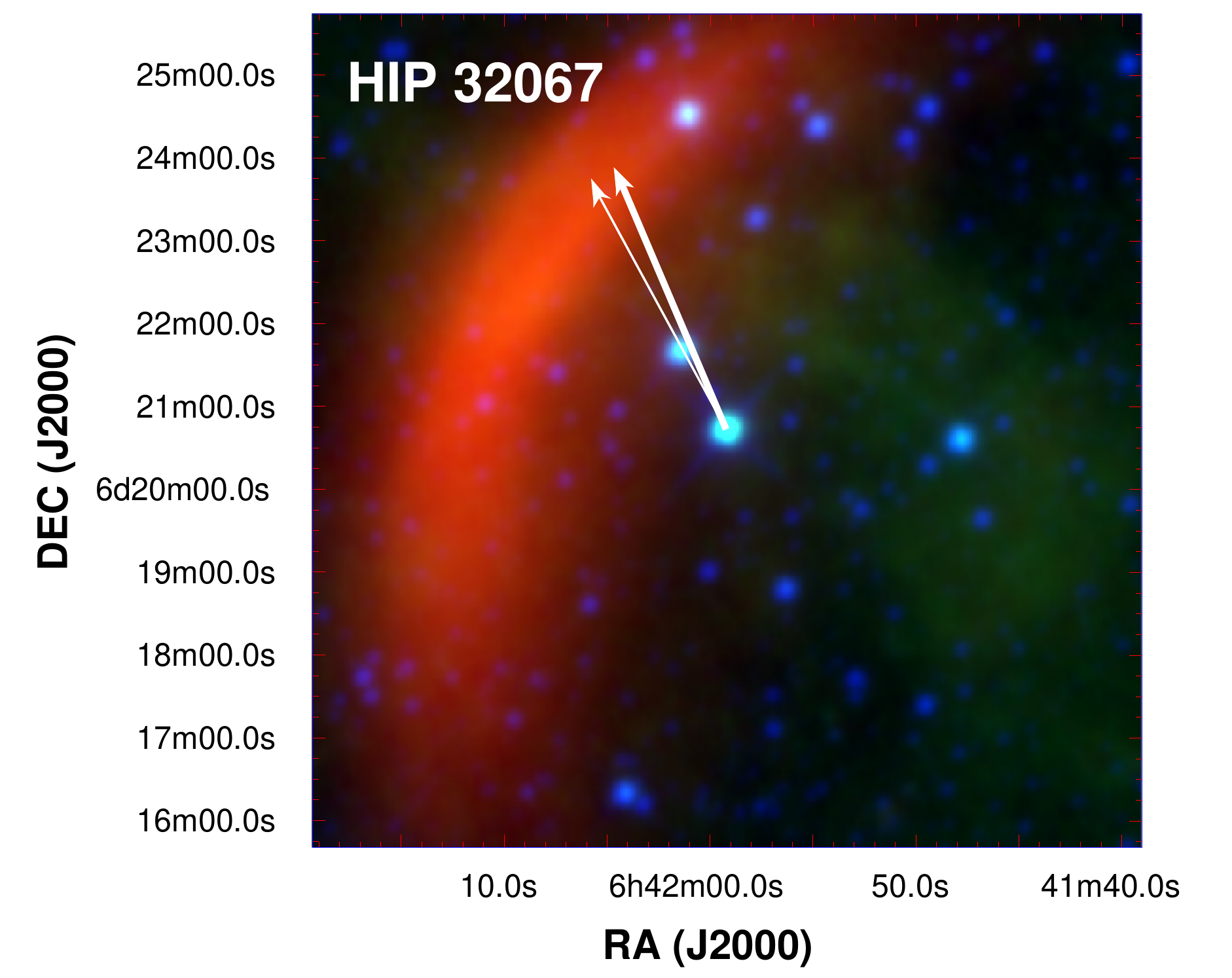}
\hfill
\includegraphics[width=0.463\textwidth,height=6.8cm]{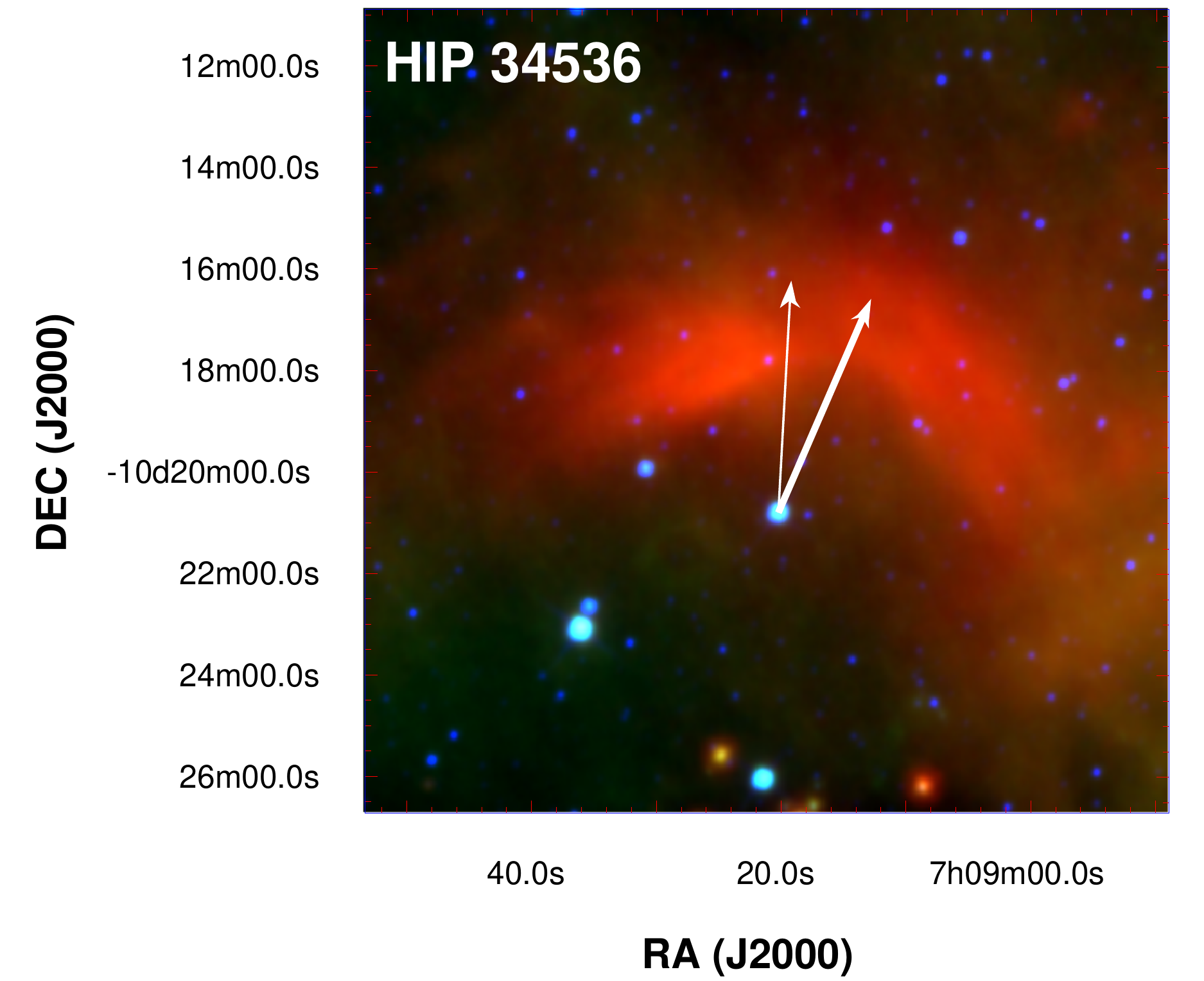}
\end{minipage} 

\begin{minipage}{\textwidth}
\centering
\includegraphics[width=0.455\textwidth,height=6.7cm]{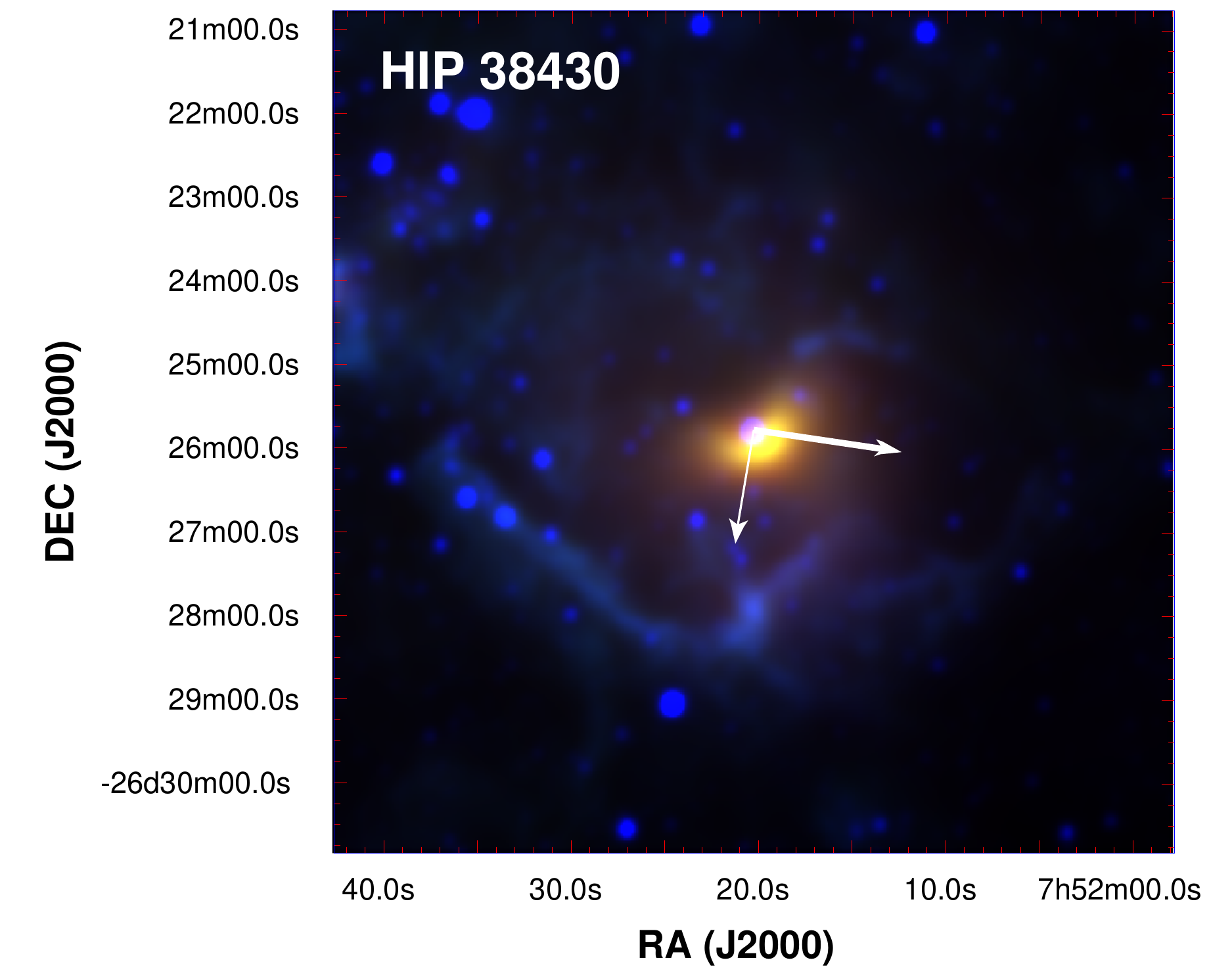}
\hfill
\includegraphics[width=0.45\textwidth,height=6.6cm]{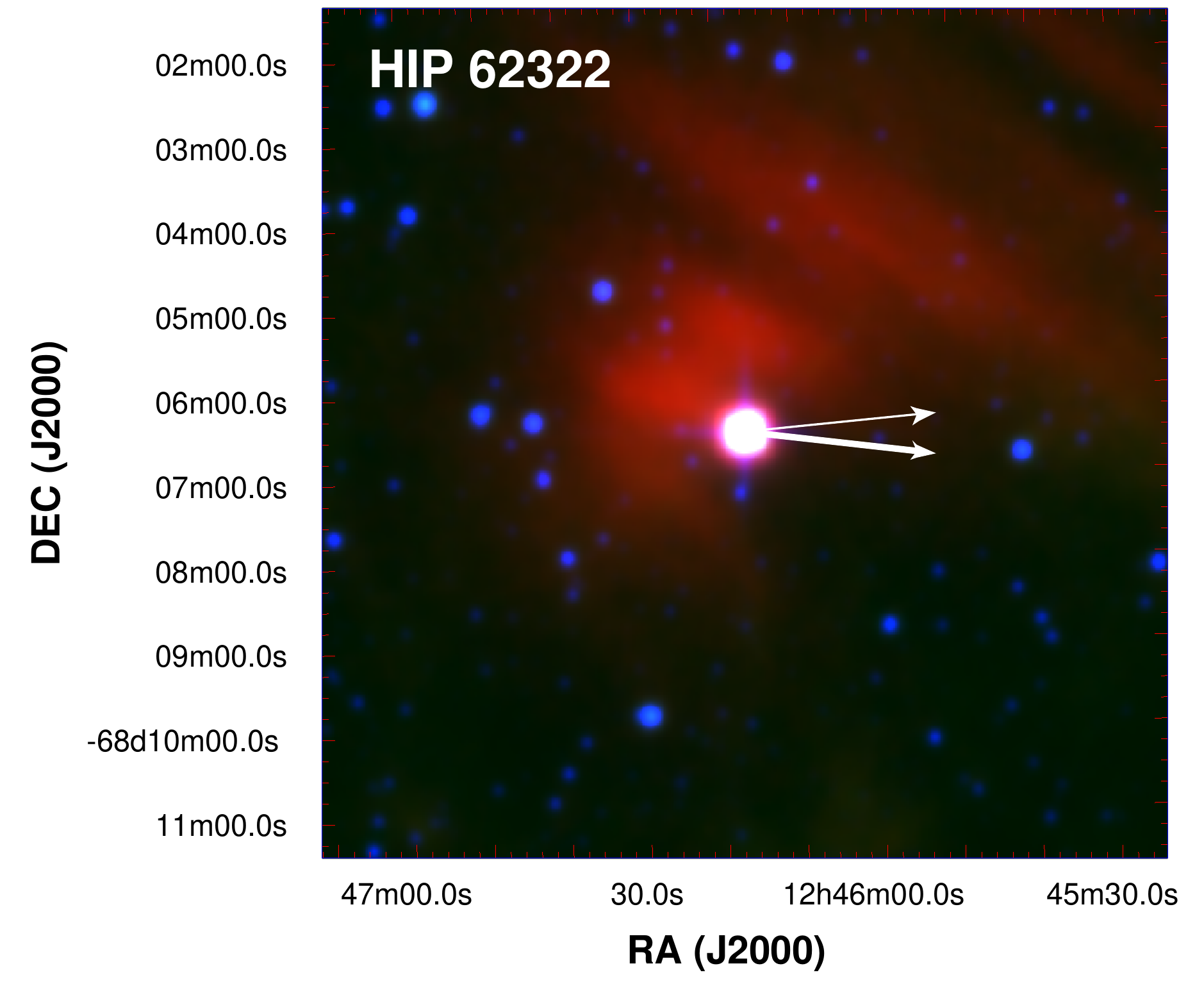}
\end{minipage}

\begin{minipage}{\textwidth}
\centering
\includegraphics[width=0.45\textwidth,height=6.7cm]{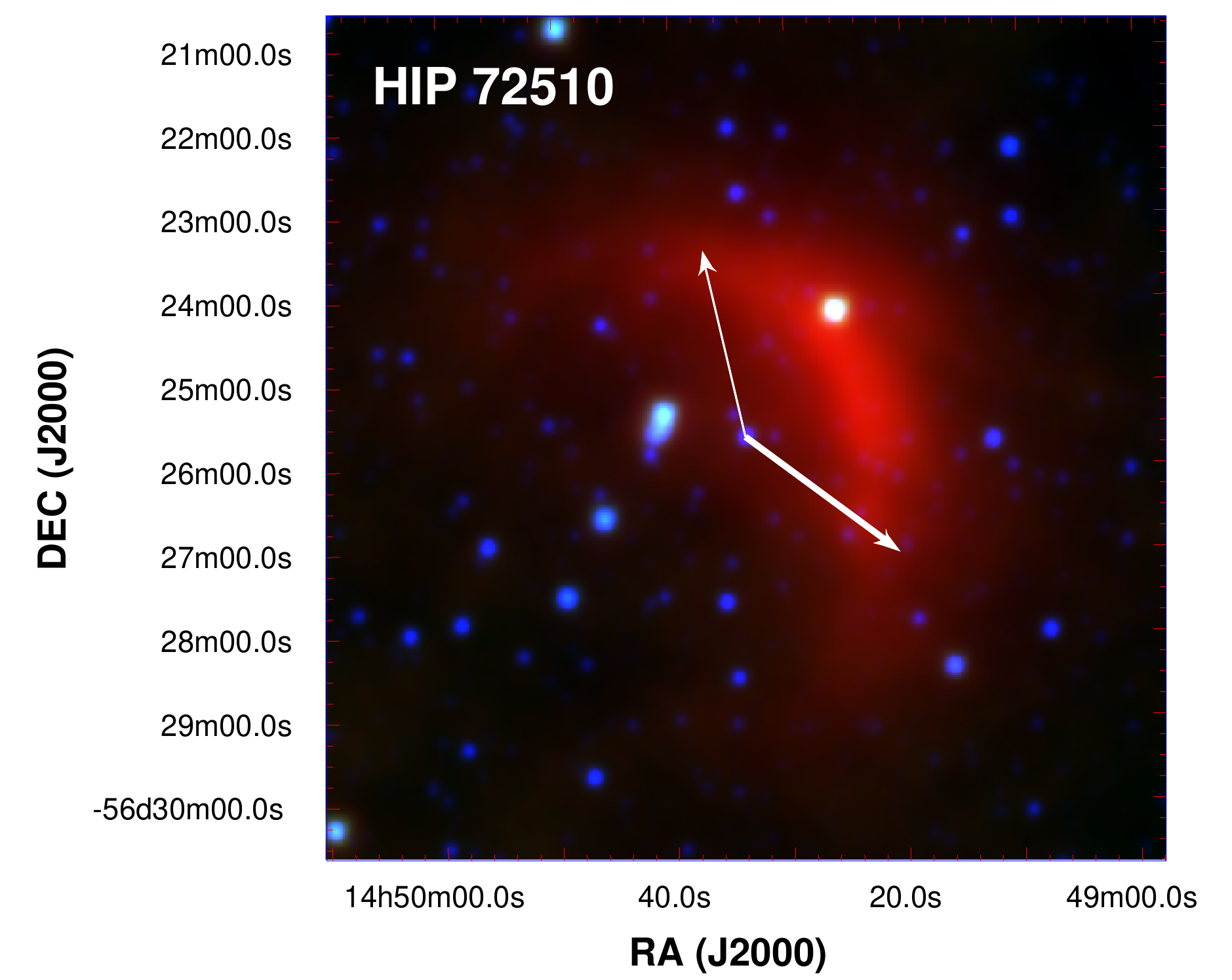}
\hfill
\includegraphics[width=0.45\textwidth,height=6.7cm]{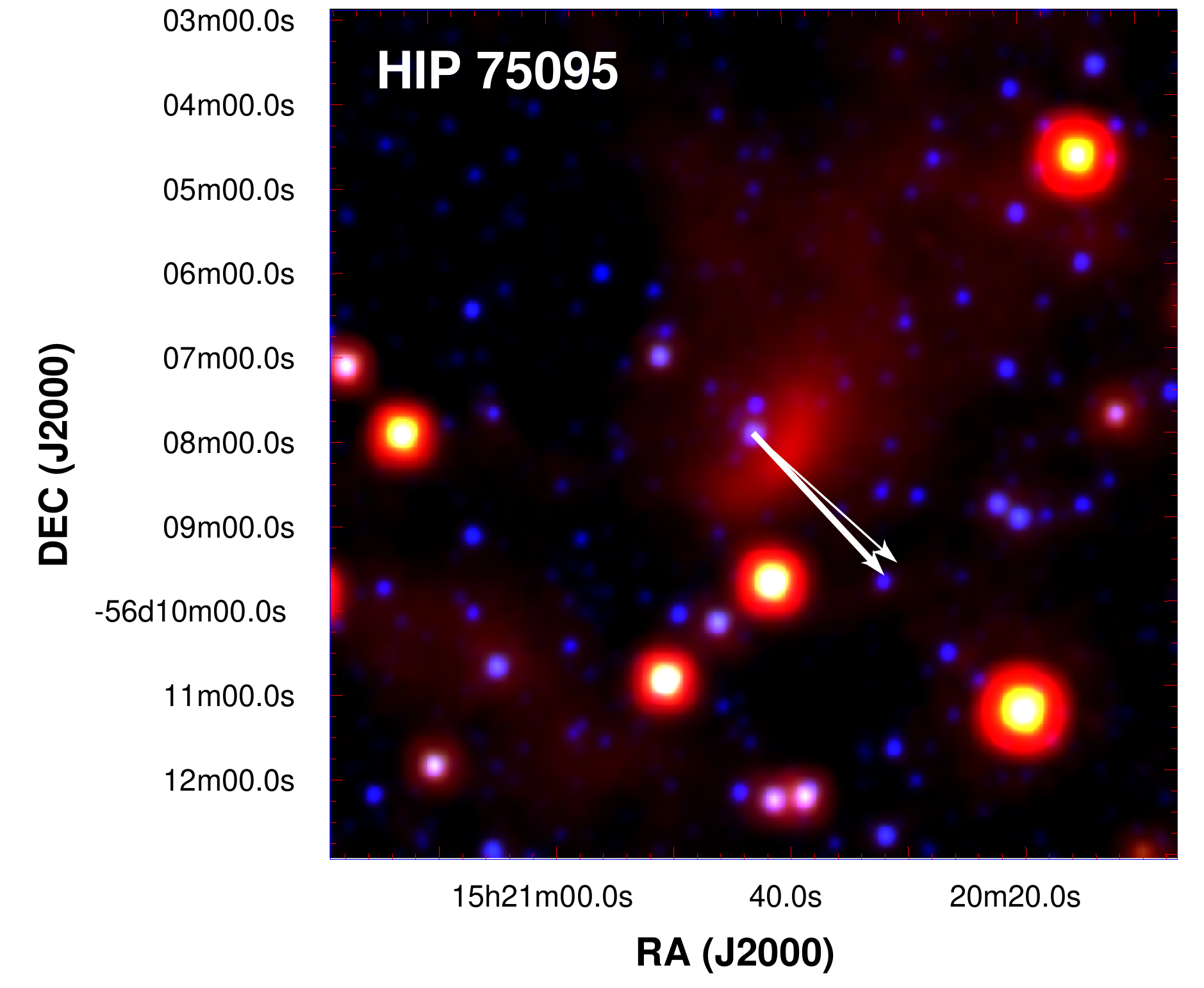}
\end{minipage}

\vspace{0.5cm}
\caption{Same as in Figs 1-2, for another six OB stars.}
\label{bwshcks13to18}
\end{figure*}

\begin{figure*}[t]
\begin{minipage}{\textwidth}
\centering
\includegraphics[width=0.47\textwidth,height=6.8cm]{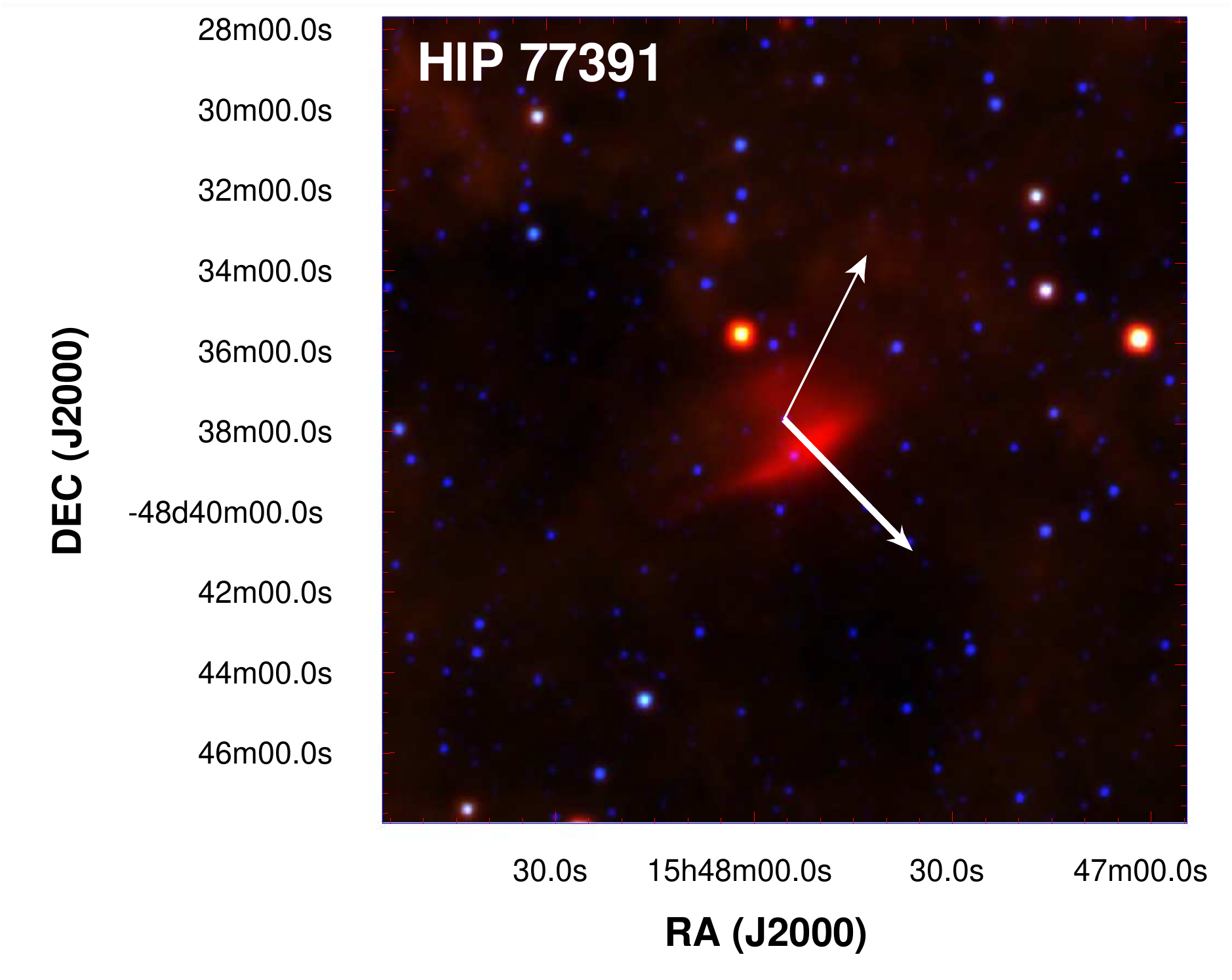}
\hfill
\includegraphics[width=0.48\textwidth,height=6.7cm]{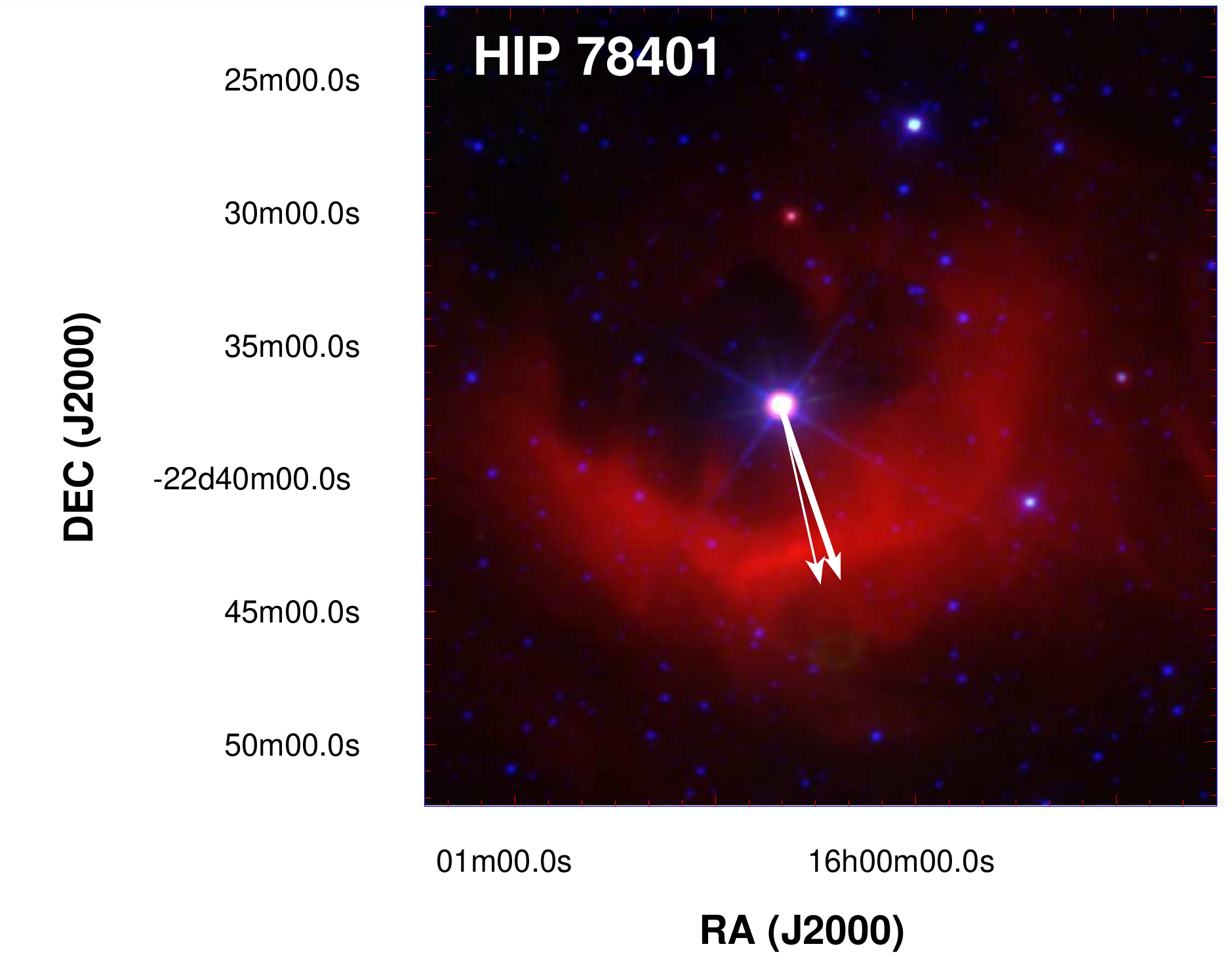}
\end{minipage}

\begin{minipage}{\textwidth}
\centering
\includegraphics[width=0.48\textwidth,height=6.9cm]{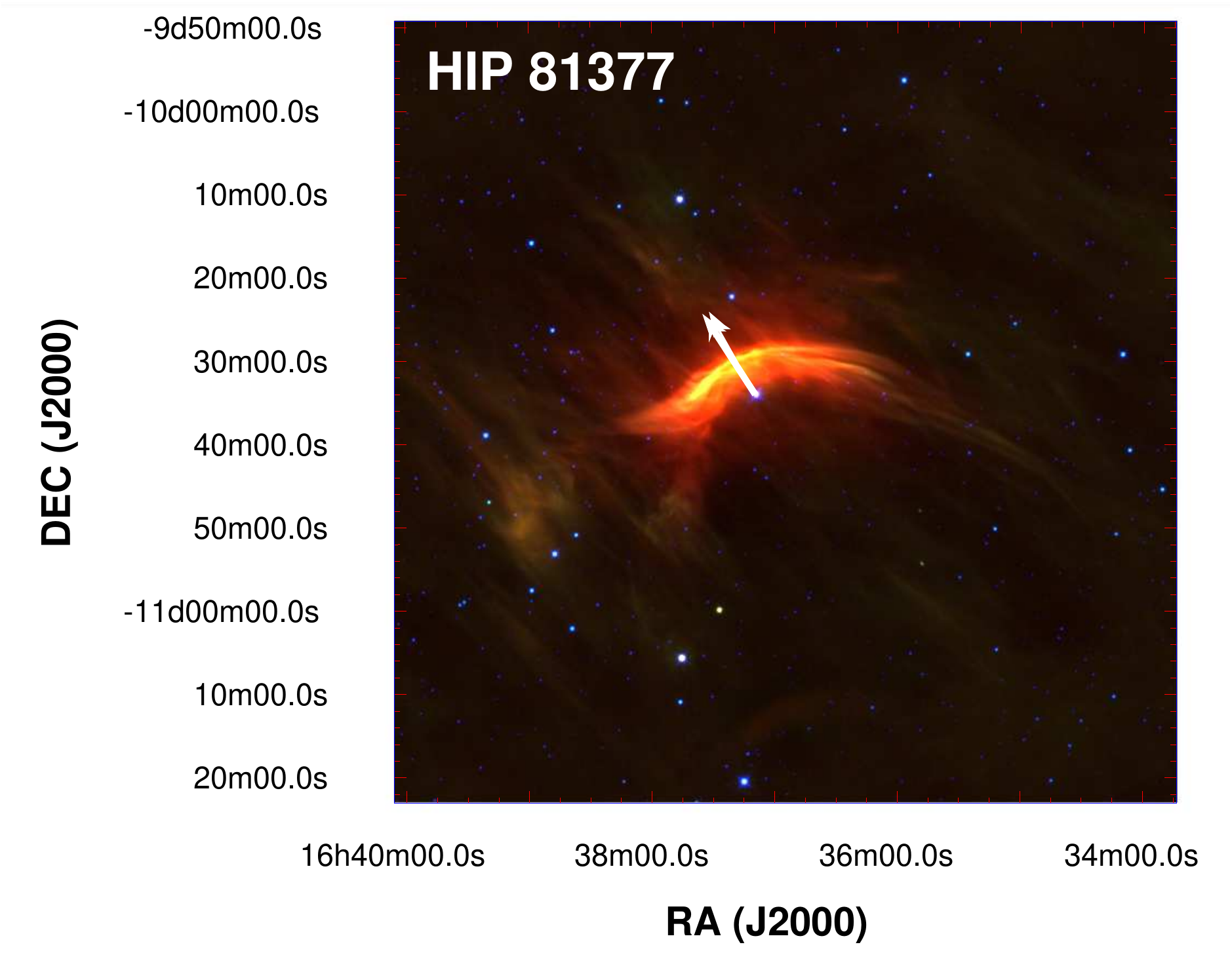}
\hfill
\includegraphics[width=0.43\textwidth,height=6.7cm]{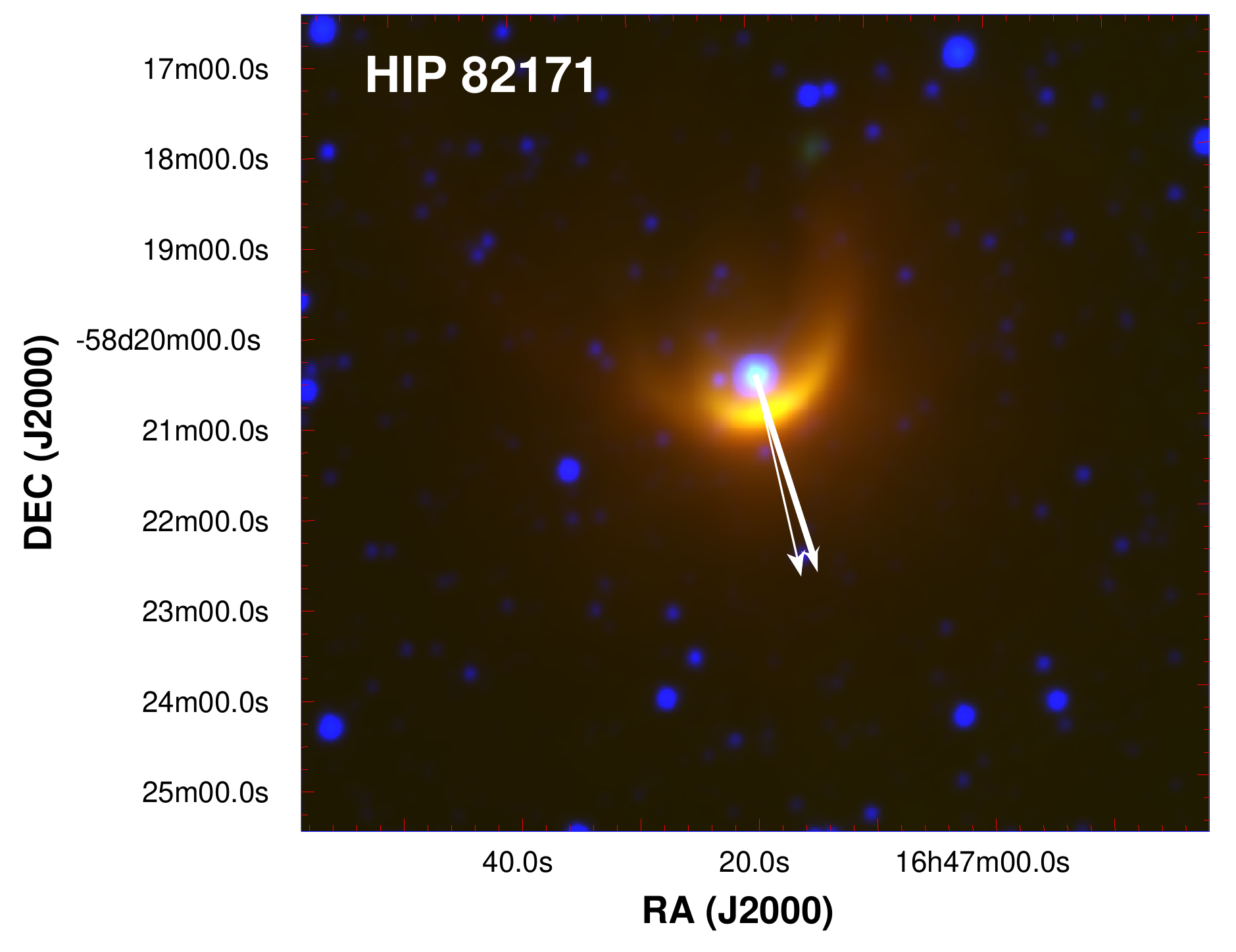}
\end{minipage}

\begin{minipage}{\textwidth}
\centering
\includegraphics[width=0.44\textwidth]{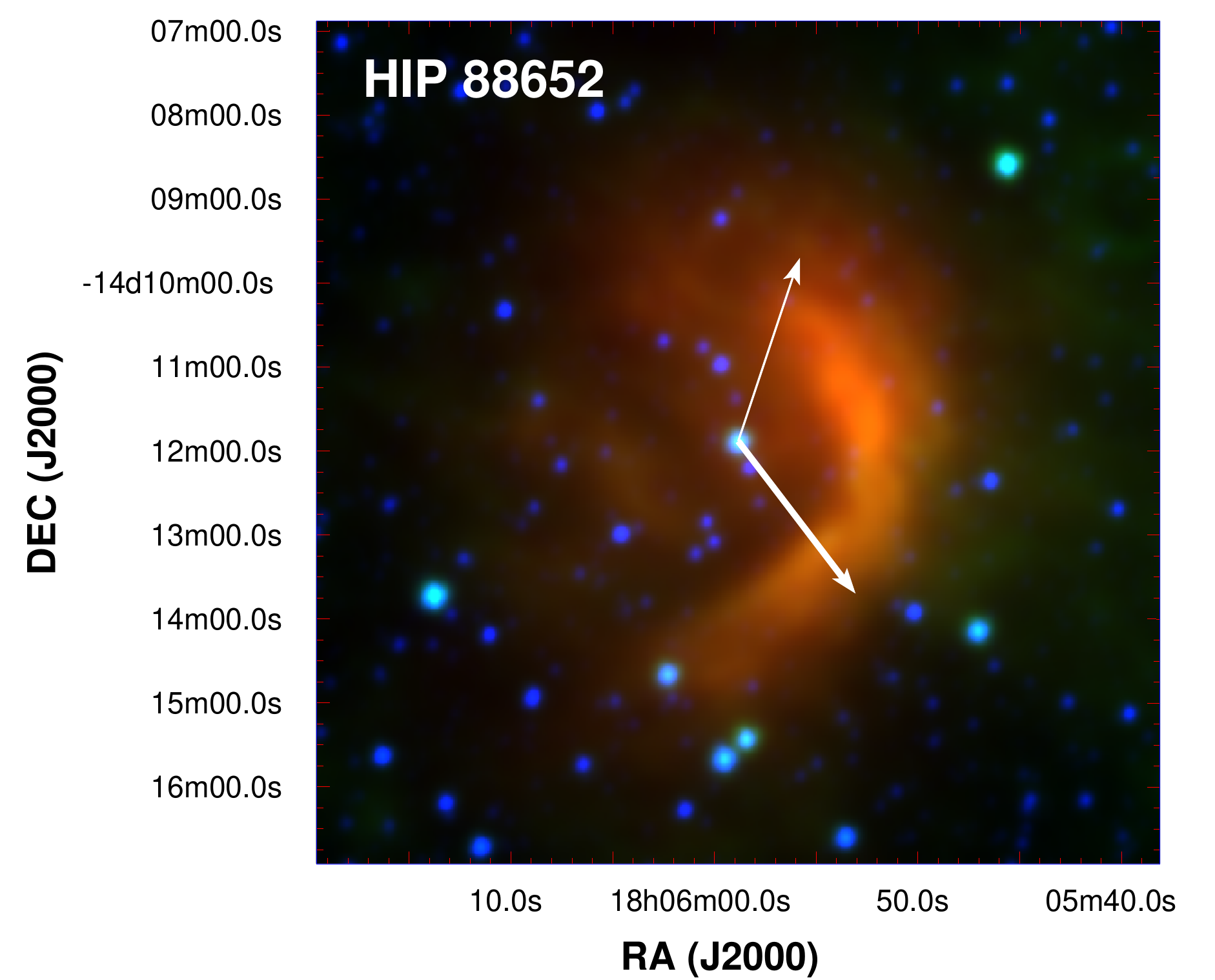}
\hfill
\includegraphics[width=0.44\textwidth]{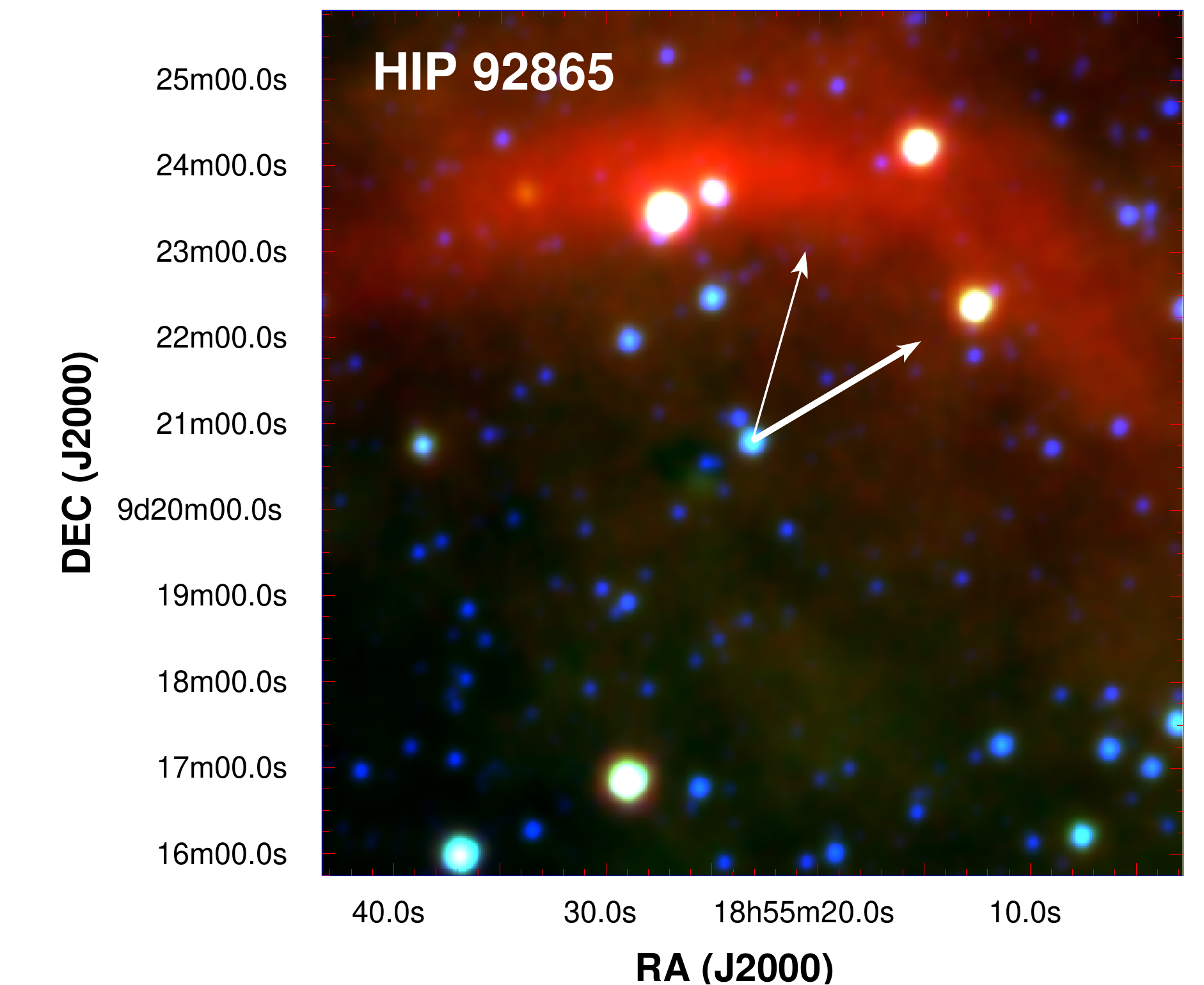}
\end{minipage}

\vspace{0.5cm}
\caption{Same as in Figs 1-3, for another six OB stars.}
\label{bwshcks19to24}
\end{figure*}

\begin{figure*}[t]

\begin{minipage}{\textwidth}
\centering
\includegraphics[width=0.46\textwidth,height=6.8cm]{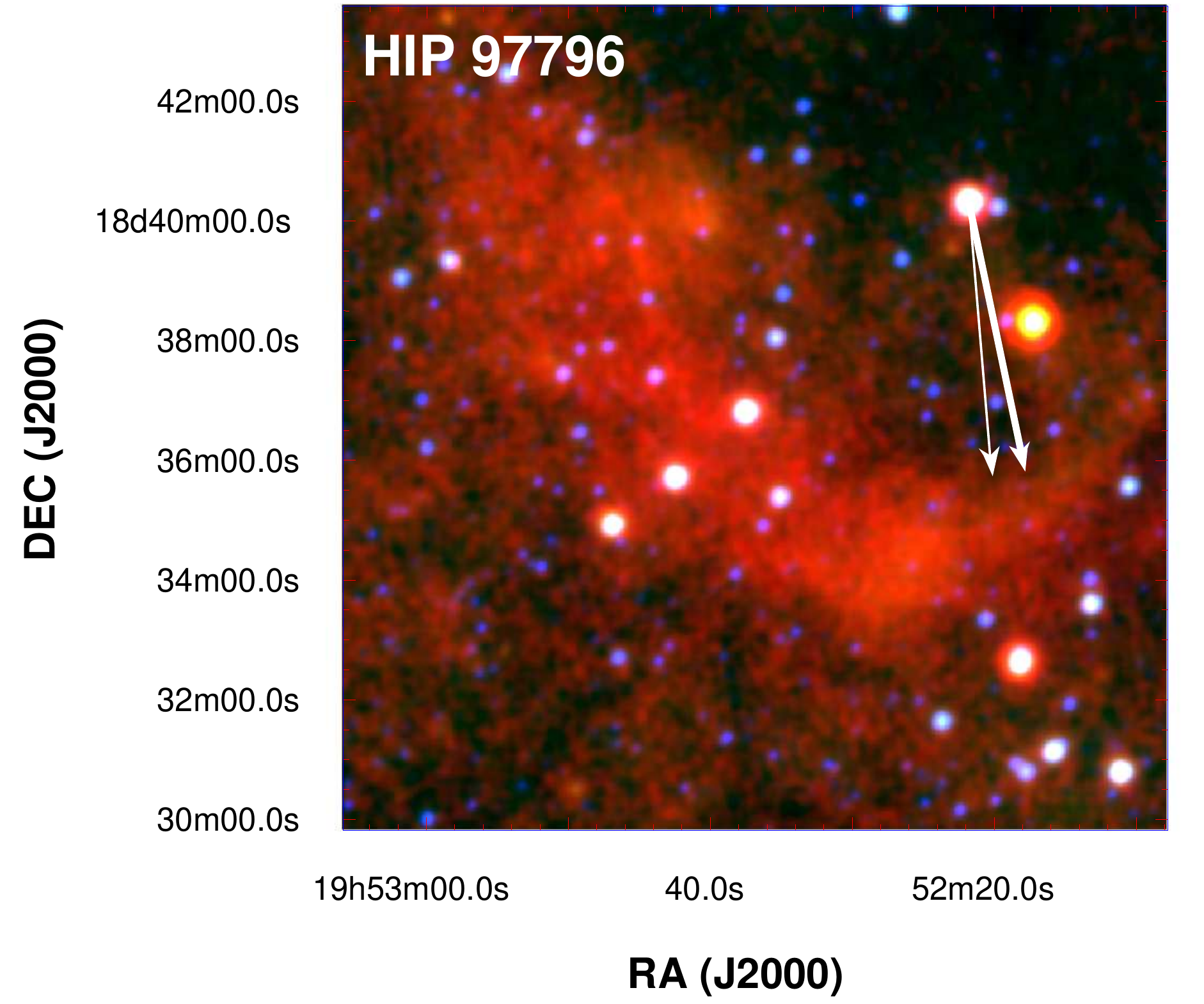}
\includegraphics[width=0.46\textwidth,height=6.6cm]{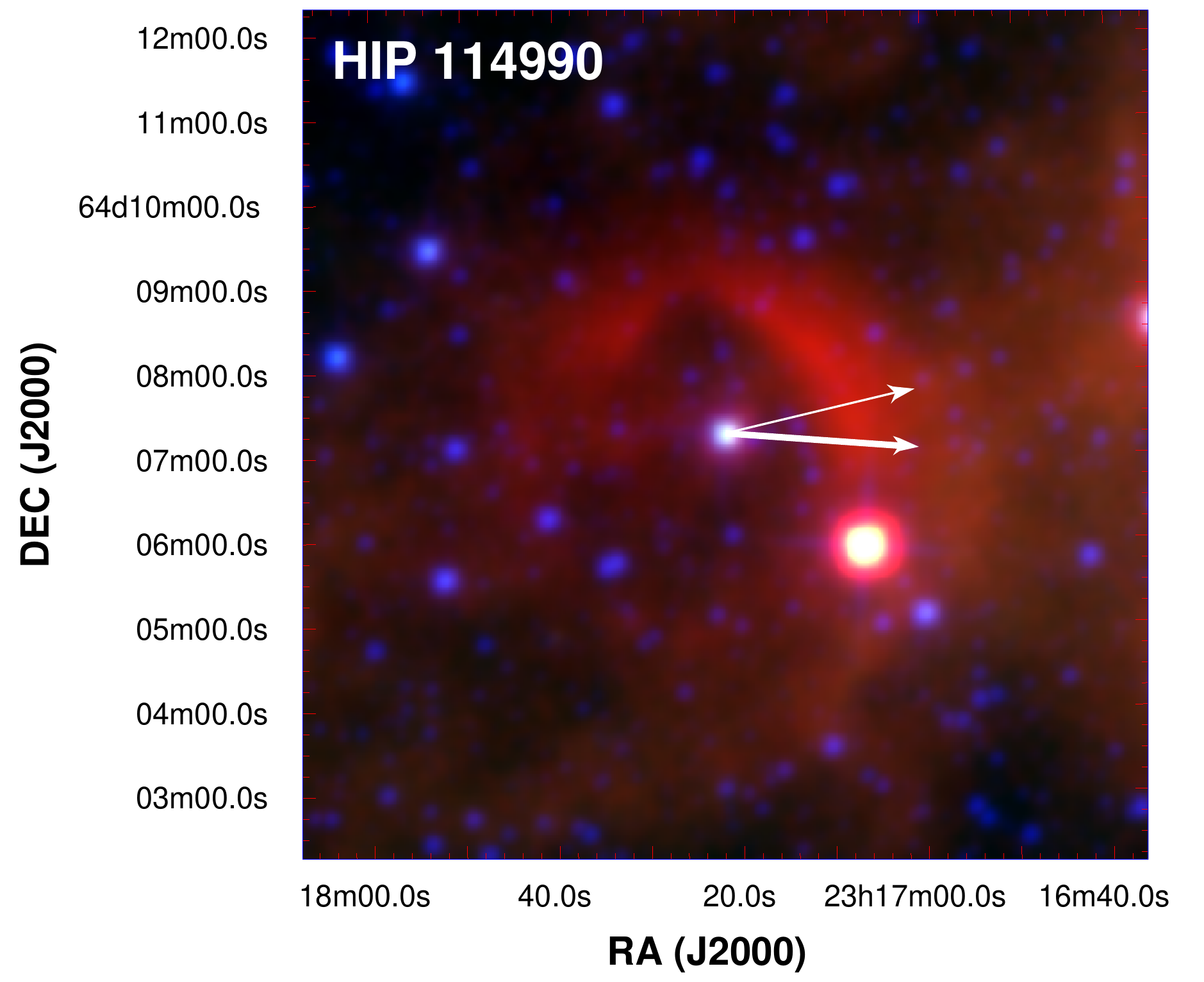}
\end{minipage} 

\begin{minipage}{\textwidth}
\centering
\includegraphics[width=0.4\textwidth]{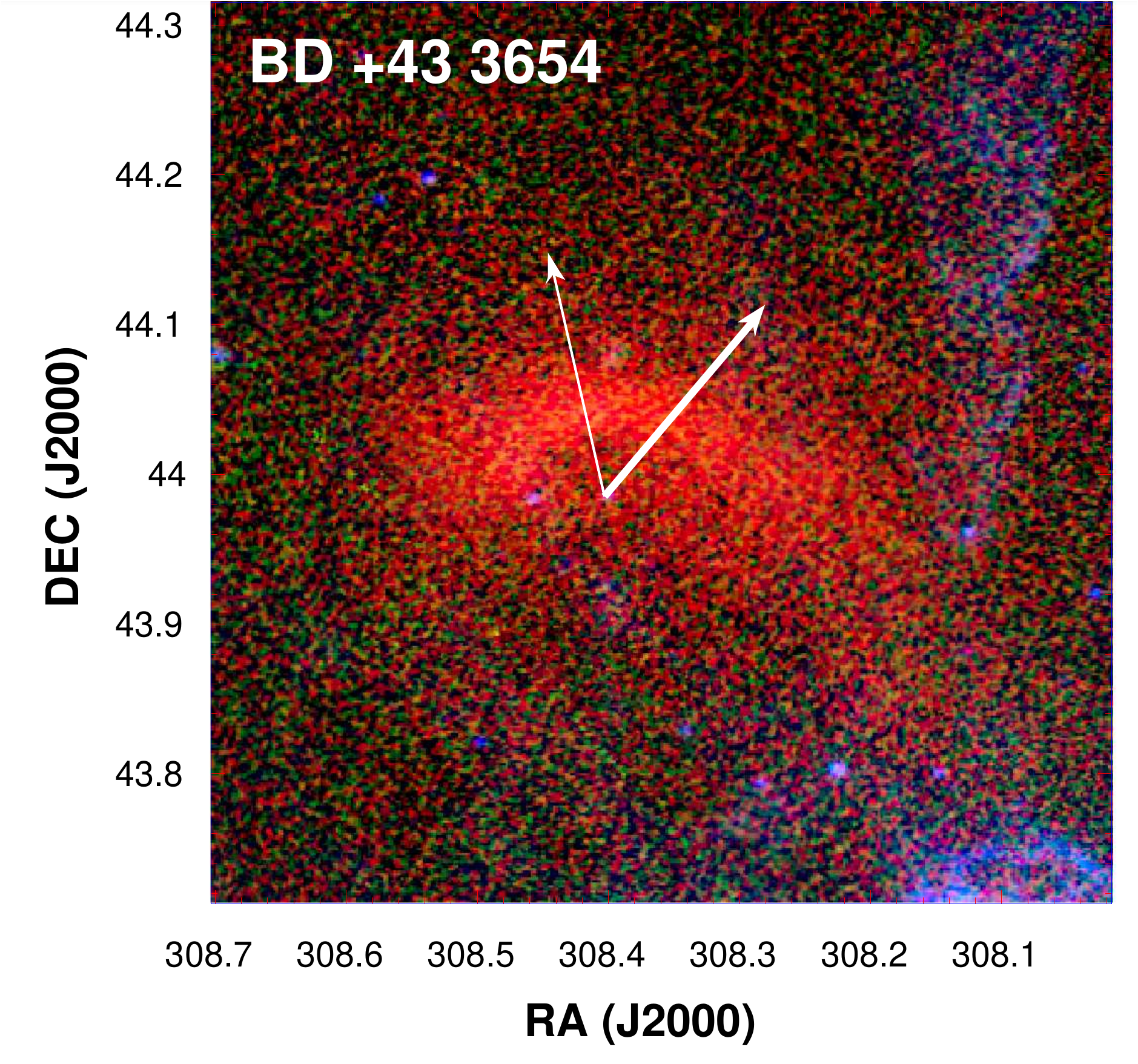}
\includegraphics[width=0.4\textwidth]{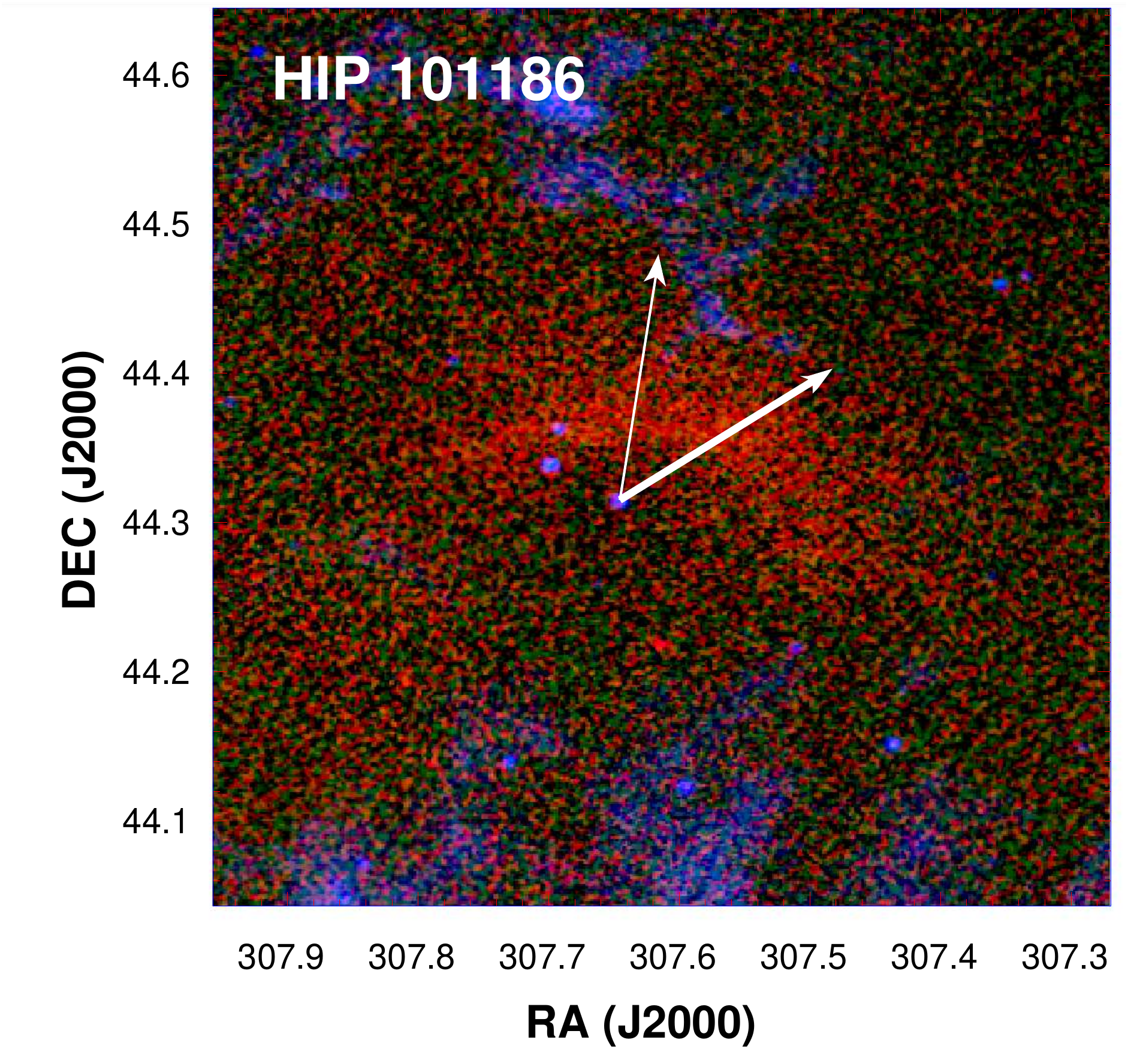}
\end{minipage}

\vspace{0.5cm}
\caption{WISE images of bow shock candidates as in Figures 1-4, 
for HIP 97796 and 114990. MSX images for BD\,+43\degr\,3654 and 
HIP 101186. Color mapping for MSX: blue=8.3 microns; green=12.1 microns;
red=21.3 microns.}
\label{bwshcks25to28}
\end{figure*}

%% file: estadistica-v2.tex
\begin{figure*}[t]
 \begin{minipage}{0.5\textwidth}
 \centering
 \includegraphics[angle=0,width=\textwidth]{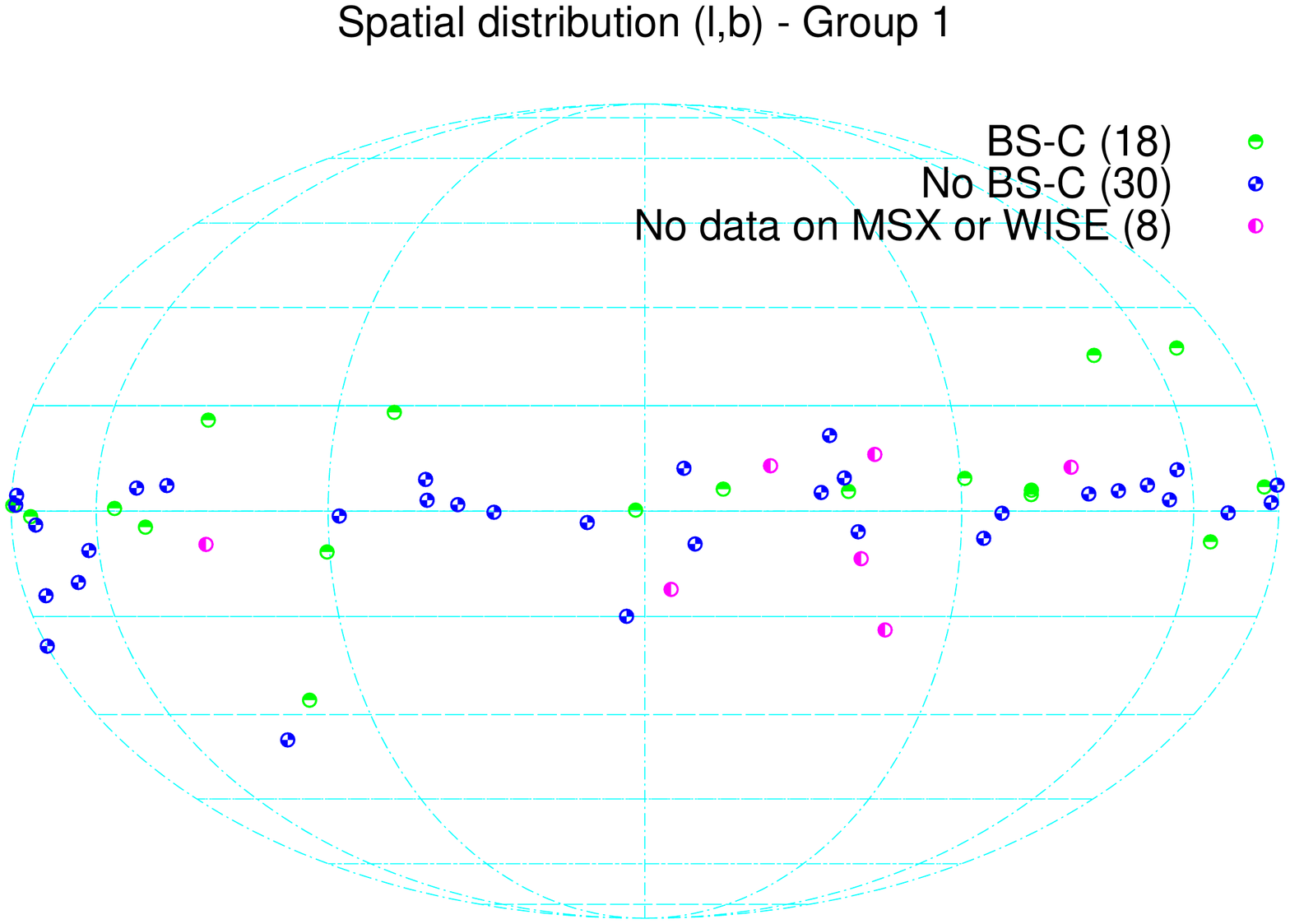}
  \caption{Distribution on the (l,b) plane of group-1 stars.}
\end{minipage} 
\begin{minipage}{0.5\textwidth}
\centering
 \includegraphics[angle=0,width=\textwidth]{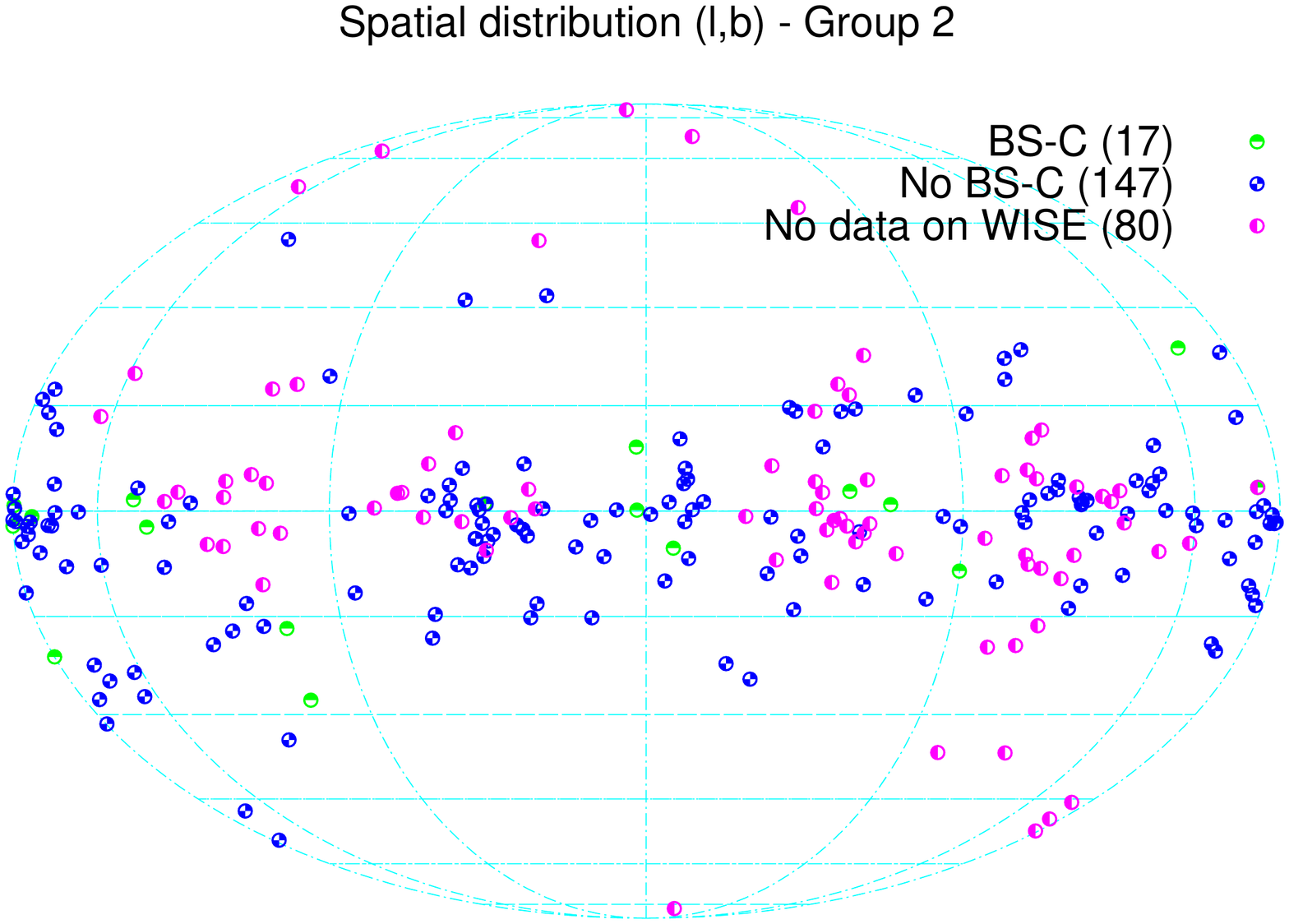}
  \caption{Distribution on the (l,b) plane of group-2 stars.}
\end{minipage}
\end{figure*}

\begin{figure*}[t]
\begin{minipage}{0.5\textwidth}
\centering
\includegraphics[angle=0,width=\textwidth]{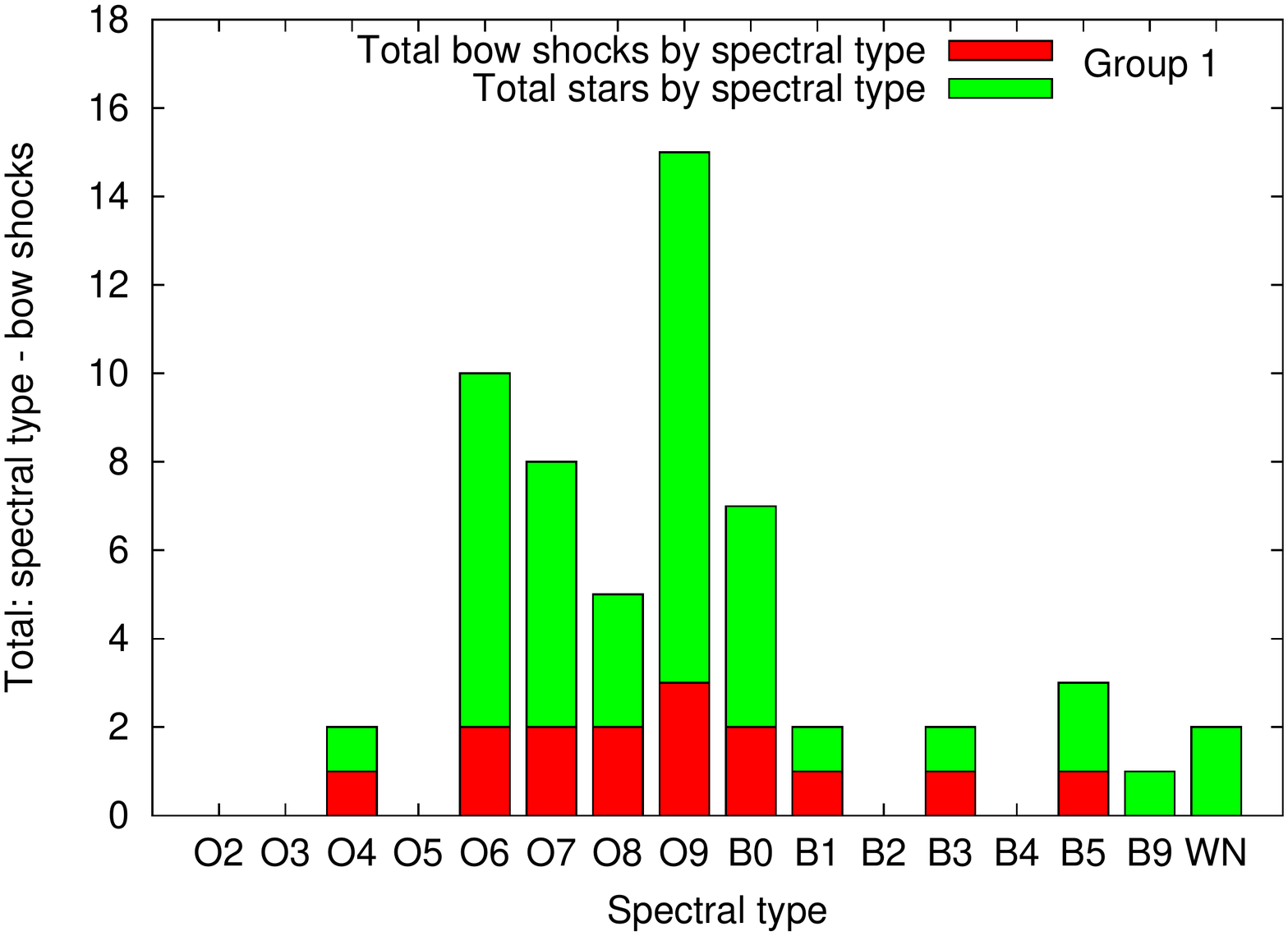}
\caption{Distribution of group-1 stars by spectral types.}
\end{minipage} 
\begin{minipage}{0.5\textwidth}
\centering
\includegraphics[angle=0,width=\textwidth]{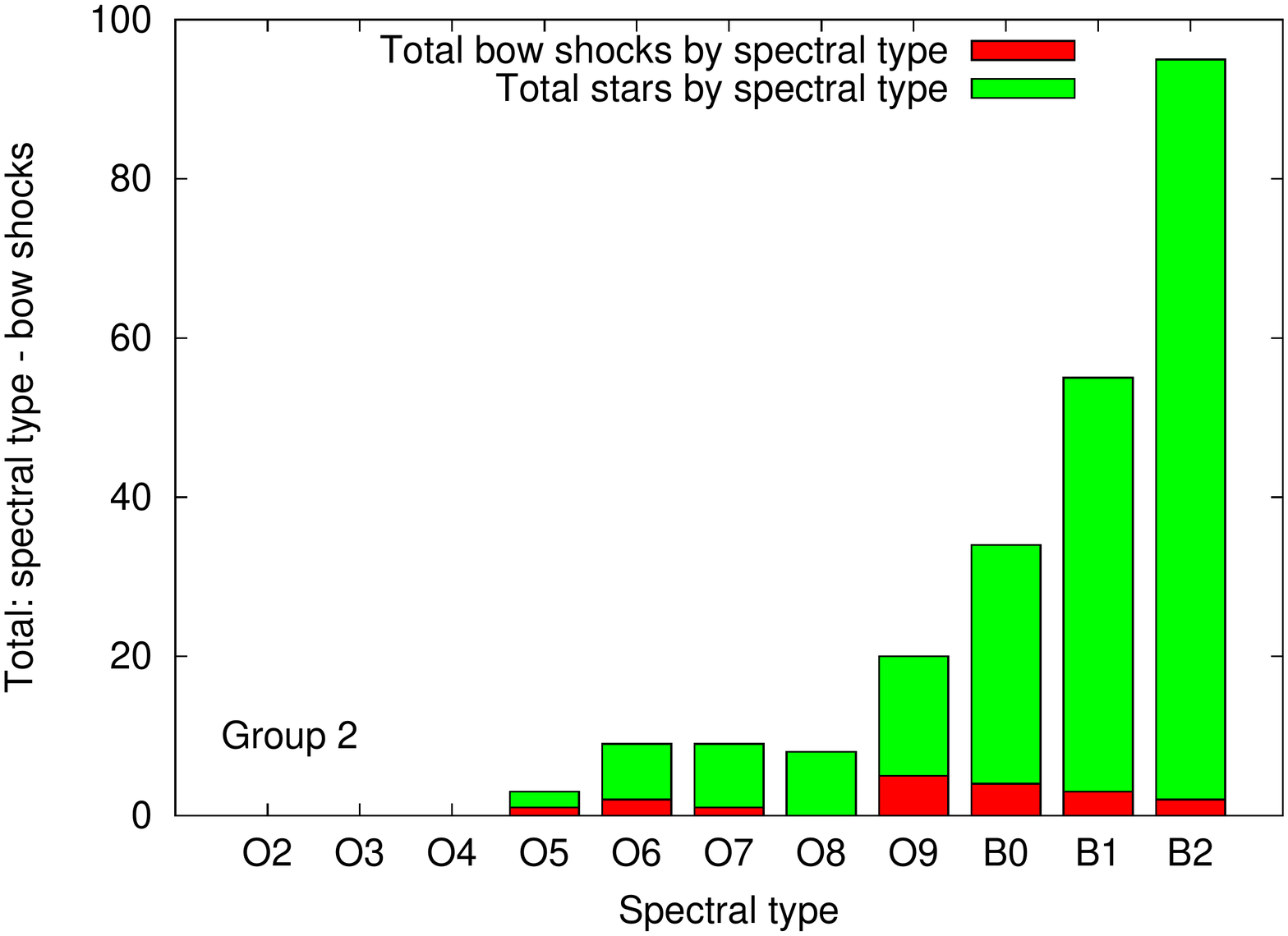}
\caption{Distribution of group-2 stars by spectral types.}
\end{minipage} 
\end{figure*}

\begin{figure*}
\begin{minipage}{0.5\textwidth}
  \centering
  \includegraphics[angle=0,width=\textwidth]{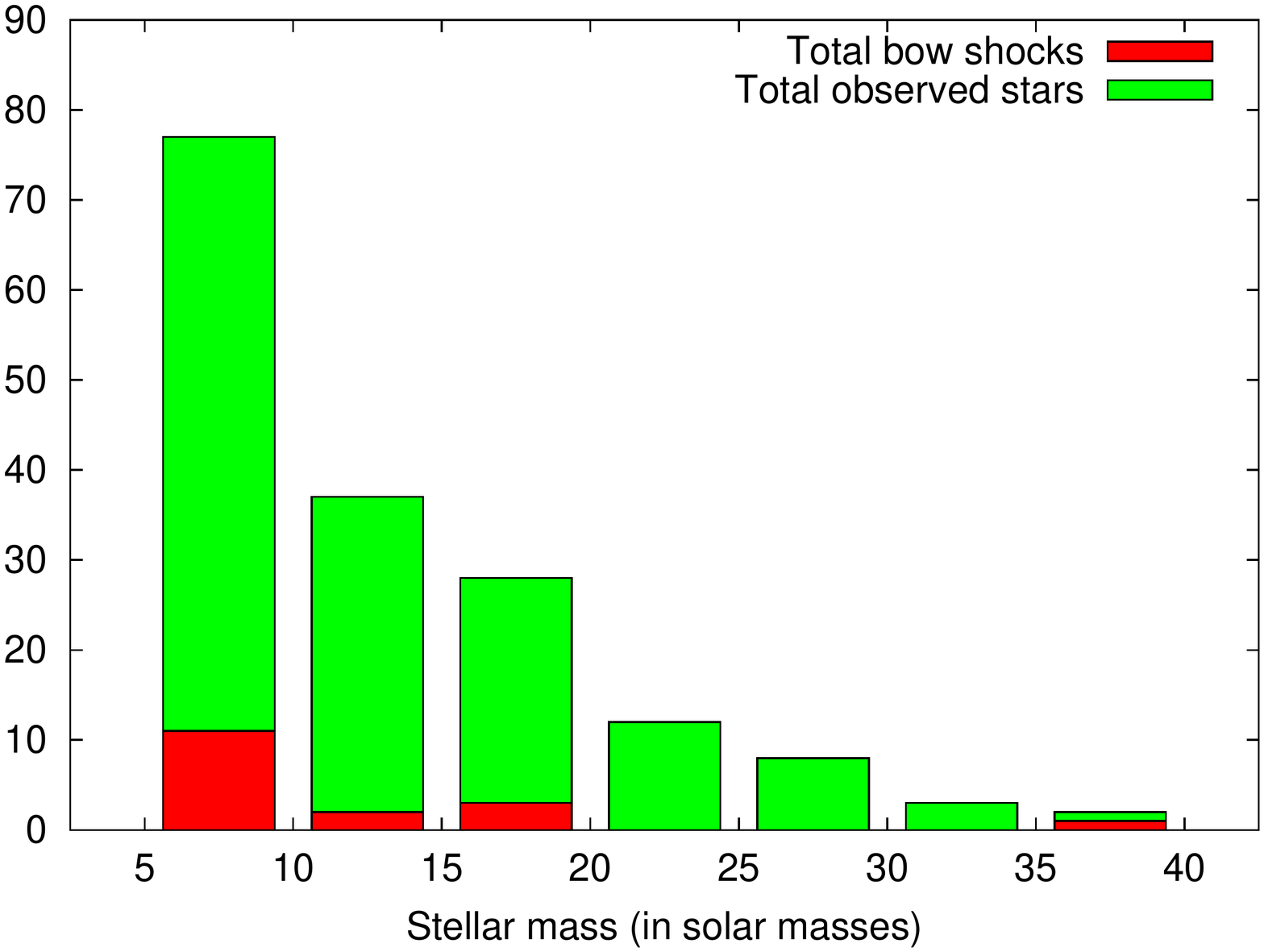}
  \caption{Distribution of group-2 stars by stellar mass (5 M$_{\odot}$ binning).}
\end{minipage} 
\begin{minipage}{0.5\textwidth}
  \centering
  \includegraphics[angle=0,width=\textwidth]{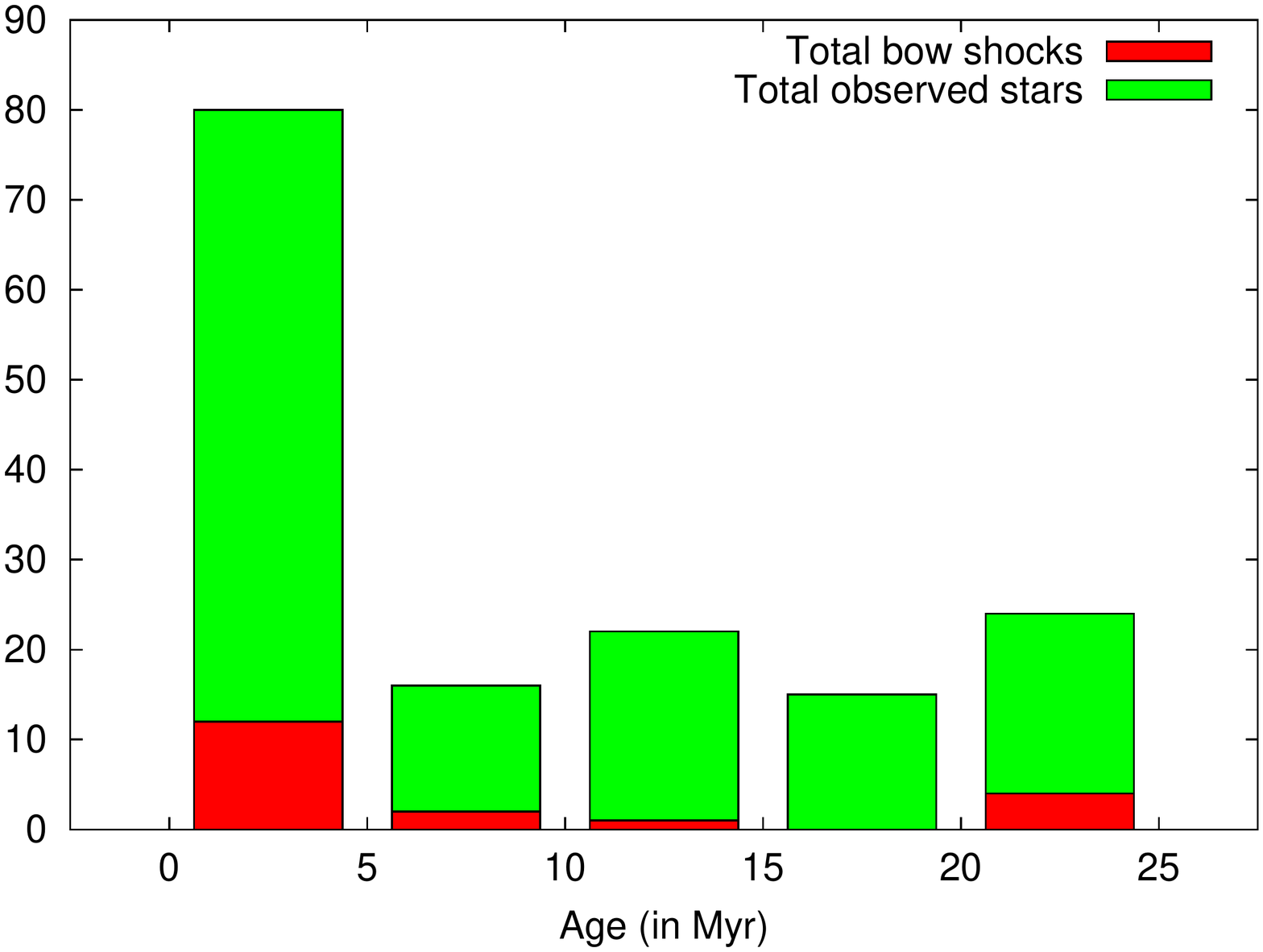}
 \caption{Distribution of group-2 stars by age (5 Myr binning).}
\end{minipage}
\end{figure*}